\documentclass[journal]{IEEEtran}
\usepackage{amsmath,amssymb,amsthm} 
\usepackage[square,comma,numbers,sort&compress]{natbib}
\usepackage{array}
\usepackage{graphicx}
\usepackage{xcolor,colortbl}
\usepackage{caption}
\usepackage{subcaption}
\usepackage{flushend}
\usepackage{ctable}
\usepackage{comment}

\newcommand{\x}[1]{\textcolor{black}{#1}}

\newcommand{\remove}[1]{}

\newtheorem{example}{Example}[]
\newtheorem{definition}{Definition}[]
\newtheorem{case}{Case}[]

\title{\x{A Framework for Automated Correctness Checking of Biochemical Protocol Realizations on Digital
Microfluidic Biochips}}

\author{Sukanta~Bhattacharjee,
		Ansuman~Banerjee,
		Krishnendu~Chakrabarty,
        and~Bhargab~B.~Bhattacharya
\thanks{A preliminary version of this paper has appeared in the proceedings of VLSI Design 2014~\cite{SukantaVLSID14}.}
\thanks{S. Bhattacharjee is with the Department of Computer Science and Engineering, Indian Institute of Technology Guwahati, Assam, India 781039. E-mail: sukantab@iitg.ac.in}
\thanks{A. Banerjee is with the Advanced Computing and Microelectronics Unit, Indian Statistical Institute, Kolkata, India 700108. E-mail: ansuman@isical.ac.in}
\thanks{K. Chakrabarty is with the Department of Electrical and Computer Engineering, Duke University, USA. E-mail:
krish@ee.duke.edu}
\thanks{B. B. Bhattacharya is with the Department of Computer Science and Engineering, National Institute of Technology Rourkela, Odisha, India 769008. E-mail: bhargab.bhatta@gmail.com}
}

\begin{document}

\maketitle

\begin{abstract}
Recent advances in digital microfluidic (DMF) technologies offer a promising platform for a wide variety of biochemical
applications, such as DNA analysis, automated drug discovery, and toxicity monitoring. For on-chip implementation of
complex bioassays, automated synthesis tools have been developed to meet the design challenges. Currently, the synthesis 
tools pass through a number of complex design steps to realize a given biochemical protocol on a target DMF 
architecture. Thus, design errors can arise during the synthesis steps. Before deploying a DMF biochip on a safety 
critical system, it is necessary to ensure that the desired biochemical protocol has been correctly implemented, i.e., 
the synthesized output (actuation sequences for the biochip) is free from any design or realization errors. We  propose  
a symbolic constraint-based analysis  framework for checking the correctness of a synthesized biochemical protocol with 
respect to the original design specification. The verification scheme based on this framework can detect several 
post-synthesis fluidic violations and realization errors in 2D-array based or pin-constrained biochips as well as in 
\x{cyberphysical systems}. It further generates diagnostic feedback for error localization. We present experimental 
results on the polymerase chain reaction (PCR) and {\em in-vitro} multiplexed bioassays to demonstrate the proposed 
verification approach.
\end{abstract}


\section{Introduction}
\noindent
Advances in digital microfluidic (DMF) technologies offer a promising platform for a variety of biochemical
applications, ranging from massively parallel DNA analysis and computational drug discovery to toxicity monitoring and
diagnosis~\cite{Jebrail}. The steady increase in complexity of biochemical protocols being implemented
on DMF biochips has motivated  the development of automated synthesis tools that take protocol descriptions as input and
systematically synthesize efficient and optimized implementations. In this fast evolving landscape, ensuring that the
synthesized biochips are functionally correct is crucial; an error in the functionality can not only delay product
deployment, but also  endanger human life in biomedical applications.

\begin{figure*}[!t]
\centering
\includegraphics[width=.8\textwidth]{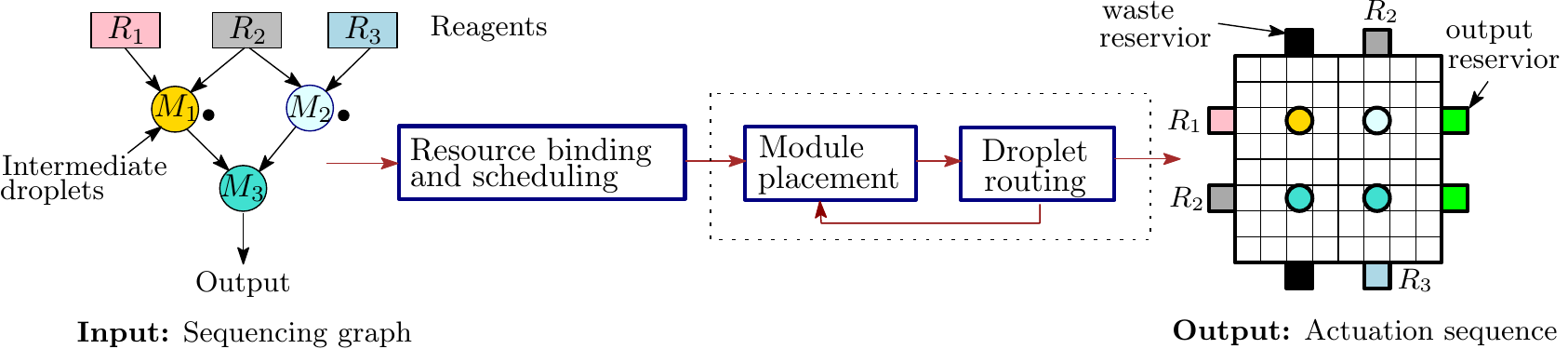}
\caption{Synthesis flow of DMF biochips}
\label{fig:synthesis}
\end{figure*}

A synthesis tool~\cite{BriskDMFBSynthesisFramework} typically executes a series of transformations for
implementing an input protocol description on a target architecture, as shown in Fig.~\ref{fig:synthesis}. At first, the
behavioral description of a bioassay (operations and dependence) is represented using a sequencing graph with
appropriate design constraints (e.g., array area, completion time, resource constraints). This is followed by a sequence
of synthesis steps, which include resource binding, scheduling, module placement, and droplet routing. Finally, a
detailed layout of the DMF biochip (DMFB) along with the sequence of actuation steps are generated. Automated
synthesis tools for DMF protocols today have reached a moderate level of sophistication and are ready to be deployed in
practice~\cite{OnePassSynthesisDAC14,GrissomOnlineFramework}.

As microfluidic technologies are poised to replace laboratory-based biochemical procedures, it is important to ensure
the reliability and accuracy of a protocol realization. This becomes even more critical when biochips are deployed in
life-critical or real-time systems. This necessitates the development of an automated procedure that can formally verify 
the synthesized implementation against the high-level design objectives, i.e., the input biochemical protocol to be 
realized and target-specific constraints. Post-synthesis verification is also important when a bioassay needs to be 
remapped onto an upgraded microfluidic platform. Verification in microelectrofluidic systems was classified as being 
either subjective or objective by Zhang  {\em et al.}~\cite{ZhangKrish}  but no mechanized procedure was proposed.  A 
preliminary attempt towards design verification of flow-based microfluidic biochips has been reported by McDaniel {\em 
et al.}~\cite{BriskCFMBVerification}.  More recently, a one-pass symbolic correct-by-construction synthesis technique 
that satisfies fluidic and layout constraints has been proposed by Keszocze  {\em et al.}~\cite{OnePassSynthesisDAC14}.

In this paper, we propose an automated correctness checking framework for verifying post-synthesis realizations of 
biochemical protocols on a target DMF architecture. We adopt a \x{symbolic constraint-based analysis method} to achieve 
our objective. \x{Given an actuation sequence generated by a synthesis tool for a given input sequencing graph to be 
realized on a DMF architecture, we adopt an incremental constraint-based analysis approach for checking if the actuation 
sequence indeed takes the assay through a safe state sequence free from fluidic constraint violations. 
We continue this checking procedure over the entire actuation sequence, as specified in the synthesized output and check 
for violations to ascertain that the generated actuation sequence is indeed safe for actual deployment. In addition, we 
reconstruct a sequencing graph as we traverse through the DMF states, and check it for operational correspondence with 
the original one to check if there are any timing or realization errors on the target architecture. Moreover, the 
verification tool notifies the user where the design rules are violated. This additional capability helps in error 
localization.}

\x{The objective of our framework is to detect synthesis tool as well as realization errors. Existing synthesis tools 
cycle through several optimization strategies and generally generate the solution in multiple passes. Given the sizes 
and complexities of today's bioassays, it is therefore, quite an arduous task to debug the synthesis tool, just by 
observing the error manifestations at a user's end. Note that in the past, similar efforts were made to formally check 
the correctness of circuit implementation in connection to logic synthesis tools~\cite{AagaardL91}. However, such 
approaches did not fare well as they were not found to be practical for large systems. Also, an optimal one-pass 
synthesis tool that may work well for all protocols is hard to achieve, as evident from the scalability limitations 
witnessed in the one-pass approach~\cite{OnePassSynthesisDAC14}. Furthermore, many decisions such as the selection of 
the wash-droplet size may be made just before implementing the assay. Similarly, post-synthesis bugs may occur if the 
time needed for homogeneous mixing on the target platform is not properly characterized. Thus, rewriting a synthesis 
tool would not easily solve the problem of post-implementation correctness checking. Hence, we adopt a post-synthesis 
checking policy, where our objective is not to generate patches for synthesis tools, but to detect the errors arising 
out of imprecise synthesis strategies and the choice of target architecture.}
\x{Our proposal can be thought of as the first step towards a verification methodology in the context of microfluidic 
computation, and we expect that the generality of our verification techniques will eventually allow us 
to apply them not only to DMF-based assay implementations, but to a wider class of microfluidic paradigms. We 
demonstrate that in addition to general-purpose biochip architectures, our method is applicable to pin-constrained chips 
and cyberphysical systems. We also present an extensive case study on several real-life bioassay descriptions.}

\x{The rest of the article is organized as follows. Section II summarizes the various constraints needed for verifying 
design errors for a general-purpose DMFB. Section III discusses the correctness issues related to pin-constrained 
DMFBs. Section IV describes the design constraint analysis and the procedure for conformance checking between the input 
sequencing graph and the synthesized graph. The architectural description of the verification engine and a software 
visualization tool {\em SimBioSys}, are described in Section V. Experimental results for two real life 
biochemical assays are presented in Section VI. The correctness checking procedure for fault-tolerant assays implemented 
in a cyberphysical scenario is described in Section VII. Finally, Section VIII concludes the article and highlights 
directions for future work.}

\section{Verification of general-purpose fully reconfigurable DMF Biochips}
\label{sec:verification_general_purpose_dmfb}
\noindent
In this section, we describe several facets of the verification problem that may arise in the context of fully
reconfigurable DMFBs. We assume that a DMFB platform is capable of performing basic fluidic operations such as
dispensing a droplet from a reservoir, moving a droplet, and mixing/splitting of two droplets. 

The dilution of a sample and the mixing of reagents are basic steps in a biochemical protocol~\cite{MitraATSP}. 
The concentration factor $(CF)$ of a sample is defined as the ratio of initial volume of the raw sample to the final 
volume of the prepared mixture. Thus, $0 \leq CF \leq 1$. 

In this work, we assume the ($1:1$) mixing model, i.e., every  mix-split cycle consists of a mix operation between two 
unit-volume fluid droplets followed by a balanced split operation of the mixed fluid. Hence, for the sake of 
convenience, the $CF$  of a reagent in a target droplet is expressed as an
$n$-bit binary fraction (by rounding it off beyond the $n$-th bit), when an accuracy level of $n$ (i.e., maximum 
allowable error in $CF$  is $\frac{1}{2^{n+1}}$) is desired in $CF$.
In other words, the $CF$ of a target droplet can be expressed as $\frac{x}{2^n}$, where $x \in \mathbb{Z}^+$, $0 \leq x
\leq 2^n -1$. Note that, the $CF$ of a raw sample and a buffer can be represented by $\frac{2^n}{2^n}$ and 
$\frac{0}{2^n}$ respectively.

The following example demonstrates the various challenges that may arise in
a protocol-verification problem.

\begin{figure*}[!t]
\centering
 \includegraphics[width=.75\textwidth, height=94pt]{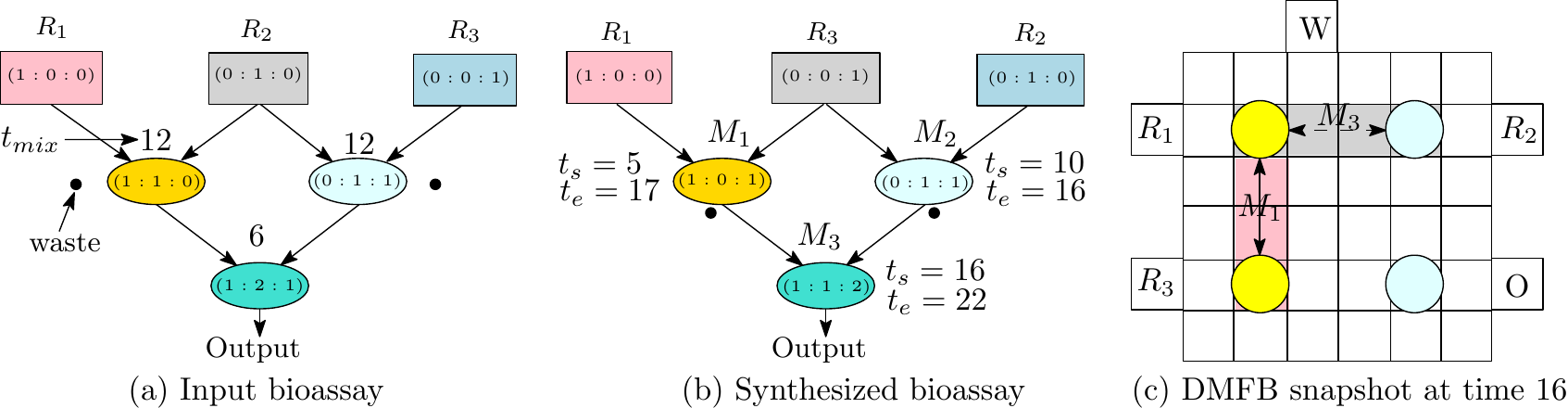}
\caption{Behavioral description of (a) input and (b) synthesized bioassay with an (c) error instance }  
\label{fig:motivating_ex}
\end{figure*}

\begin{example}
\label{ex:1}
\normalfont
Fig.~\ref{fig:motivating_ex} shows the behavioral description of an input bioassay and the corresponding one
extracted from the realization generated by the synthesis tool. The input specification is given as a sequencing
graph ($SG$)~\cite{KrishSu}, which is commonly used for describing dependencies between assay operations (e.g., 
dispensing, mixing). Fig.~\ref{fig:motivating_ex}(a) represents the input $SG$ for preparing a mixture of 
three input reagents $(R_1,R_2,R_3)$ in a ratio $(1:2:1)$, where $t_{mix}$ represents the mixing time required for 
homogeneous mixing. Our verification tool extracts a  directed acyclic graph (DAG) representing the realized bioassay. 
Fig.~\ref{fig:motivating_ex}(b) shows the $SG$ of the synthesized bioassay, where $t_s$ and $t_e$ denote the 
mixing start and end times respectively.  Unfortunately, the actuation sequences generated by the synthesis tool 
incorrectly interchanges $R_2$ and $R_3$.  As a result, the output droplets produced in Fig.~\ref{fig:motivating_ex}(b) 
have a different ratio of input reagents $(1:1:2)$, i.e., the input bioassay is realized incorrectly. This is an
instance of the error $e_7$ in Table~\ref{tab:design-error}. From Fig.~\ref{fig:motivating_ex}(b),
it can be observed that the mixing time of $R_2$ and $R_3$ is 6, but the input bioassay specifies that $R_2$ and $R_3$
should be mixed for 12 time units. Hence, an inhomogeneous mixing might have taken place $(e_6)$, resulting in a 
violation.

A different design error can be observed at the output node in Fig.~\ref{fig:motivating_ex}(b), where mixing starts
at 16. \x{In the snapshot of the DMFB at time 16 (Fig. ~\ref{fig:motivating_ex}(c)), mixer $M_3$ is instantiated by
receiving a droplet from mixer $M_1$. However, the latter completes the mix-operation at time instant 17 (Fig. 
~\ref{fig:motivating_ex}(b)), leading to an incorrect realization of the bioassay $(e_4)$.} \hfill $\blacksquare $
\end{example}

\ctable[
caption = Possible design errors and their consequences,
label = tab:design-error,
pos = !t,
width = 0.48\textwidth,
framesep=0pt,
framerule = 1pt,
doinside = \scriptsize
]{p{1pt}p{123pt}X}{
}{
\FL
\multicolumn{2}{c}{\textbf{Design error}} &
\multicolumn{1}{c}{\textbf{Consequence}} \ML
$e_1$ & Violation of static fluidic constraints (FC) & Unintentional mix of droplets \NN
$e_2$ & Violation of dynamic FC & Unintentional mix of droplets  \NN
$e_3$ & Dispensing to/from wrong reservoir & Incorrect fluidic operation \NN
$e_4$ & Move droplets to/from active mixers; late arrival or early exit  & Incorrect fluidic operation \NN
$e_5$ & Instantiate mixer with incorrect number of droplets & Droplet routing error or Incorrect fluidic operation \NN
$e_6$ & Mixing droplets for lesser amount of time than specified & Inhomogeneous mixing \NN
$e_7$ & Wrong mix-split operations & Incorrect realization of input assay 
\LL
}

\noindent
There can be several other design realization errors that may creep into the design process.
Table~\ref{tab:design-error} shows some possible design realization errors, which may lead to undesirable violations,
including those described in Example~\ref{ex:1}. \x{The list of errors have been compiled based on those reported in 
practice~\cite{SuRoutingDATE06, RapidDropletMixer, LuoCyberPhisicalErrorRecovery,cryoprotectiveMixtureLoC11}. We have 
also witnessed similar errors while using the synthesis tools on several common protocols, e.g., PCR.}

We have adopted a symbolic constraint-based incremental analysis framework for ensuring post-synthesis correctness of 
the DMF protocol realizations. The constraints may either be fluidic (both static and dynamic), as 
imposed by DMF design guidelines~\cite{KrishSu}, or imposed by the target architecture on which the input protocol is to 
be realized.  In doing so, for the sake of uniformity and portability across synthesis tools, we work on a simple 
intermediate instruction-set architecture language for representing and  manipulating actuation sequences (obtained by a 
simple transformation on the actuation sequences produced by a standard synthesis tool), in which the synthesized 
output obtained from a DMF synthesizer can be concisely described. Our method has two main steps. In the first step, 
we perform an incremental analysis of the fluidic operation actuation sequences generated by the synthesis tool and 
look for DMFB  realization errors. If any violation is found, the potential cause of violation, i.e., which fluidic 
operation causes violation of a design rule, is  reported immediately. At the end of the design verification process, if 
no errors are reported, the synthesized $SG$ is generated. In the second step, the synthesized sequencing 
graph is checked for conformance with the input protocol graph (Section~\ref{subsec:eqiv}). This is done by a variant 
of the labeled DAG traversal algorithm. If the synthesized $SG$ does not correctly realize the input 
specification, potential violations are reported for diagnostics.

\subsection{Modeling formalisms}
\begin{figure*}[!t]
\centering
\includegraphics[width=.8\textwidth]{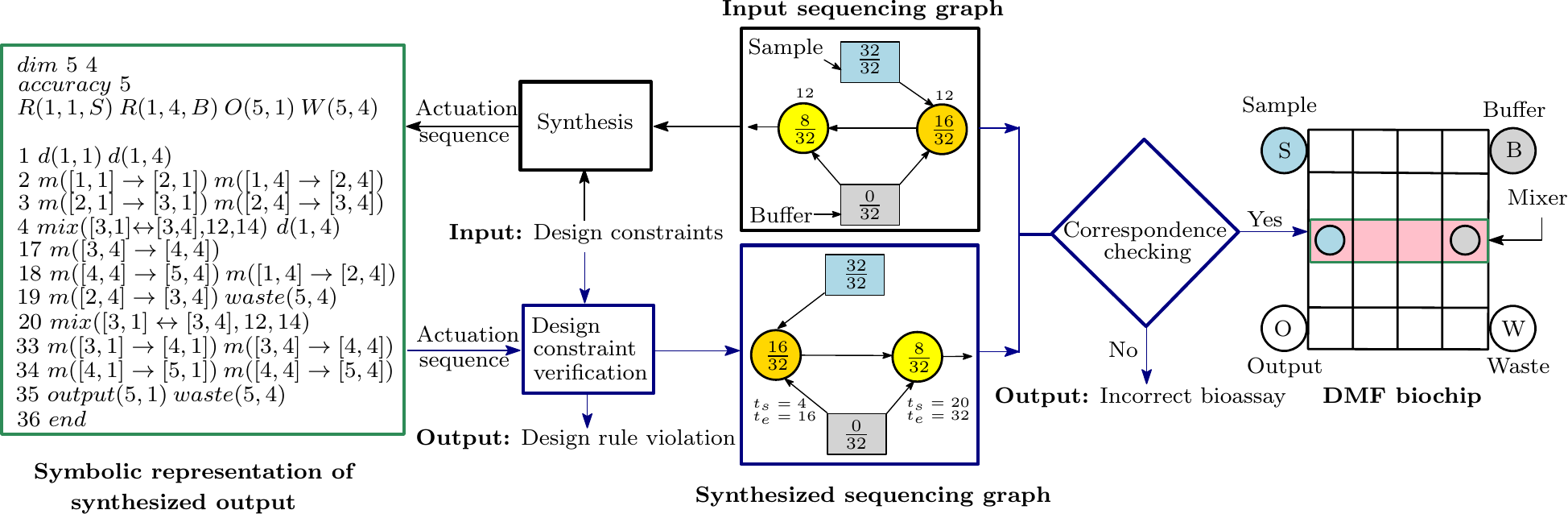}
\caption{Verification engine}
\label{fig:synopsis}
\end{figure*}

\begin{figure*}[!t]
\centering
\includegraphics[width=.8\textwidth,height=140pt]{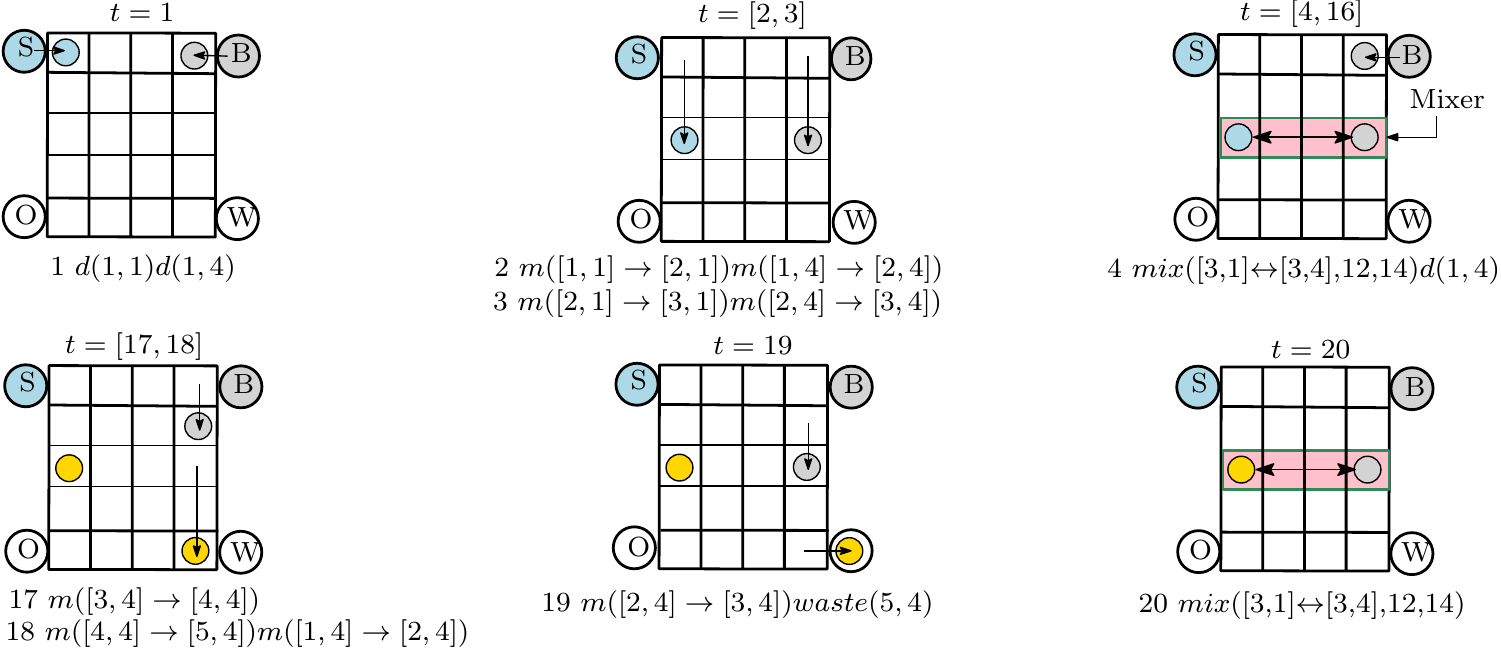}
\caption{Execution of synthesized instructions}
\label{fig:example}
\end{figure*}

\noindent
In this subsection, we address the correctness checking problem in the context of fully reconfigurable DMFBs.
The overall architecture of our verification engine is shown in Fig.~\ref{fig:synopsis}. It internally translates the 
synthesized output into an intermediate format and incrementally looks for constraint violations. If no violations are 
observed, it outputs a transformed synthesized $SG$ that is checked for conformance against the original input. 
Formally, the correctness checking problem can be  formulated as below:

\small
\begin{itemize}
\item \textit{Inputs:}
\begin{itemize}
\item Behavioral description of the protocol;
\item The output of a synthesis tool (actuation sequences to be executed on the target DMF biochip);
\item Parameters of the target DMF architecture:
		\begin{itemize}
		\item Biochip dimensions.
		\item Reservoirs and their locations.
		\end{itemize}
\item Design constraints
		\begin{itemize}
		\item Fluidic constraints (static and dynamic).
		\item Maximum bioassay completion time $(T_{max})$.
		\end{itemize}
\end{itemize}
\item \textit{Output:}
\begin{itemize}
\item Yes, if all design rules are met and the input protocol is correctly realized.
\item No, with a counterexample showing possible causes of violation.
\end{itemize}
\end{itemize}

\normalsize

\subsection*{Behavioral description of the protocol}
\noindent
The input behavioral description of the bioassay is represented as a $SG$, defined as follows. A $SG$ is a directed 
acyclic graph $G(V, E)$ with vertex set $V = \{v_i: i = 0, 1,\ldots, k\}$ and an edge set $E =
\{(v_i, v_j)$: $i, j \in \{ 0, 1,\ldots, k\}\}$. Each vertex corresponds to an assay operation (e.g., dispensing, 
mixing), while the edges capture the precedence relationships between the operations. \x{Note that for compact 
representation, multiple dispense operations of the same reagent are merged into a single node (drawn as square) in the 
$SG$. A vertex $v_i$ can optionally be labeled with a number $t(v_i)$, which denotes the time taken for 
the operation used at $v_i$.}

\begin{example}
\normalfont
The input $SG$ in Fig.~\ref{fig:synopsis} shows the dilution graph generating a target dilution having
concentration factor $\frac{8}{32}$ using only sample $(\frac{32}{32})$ and buffer $(\frac{0}{32})$ using
{\em twoWayMix}~\cite{MitraATSP} method. It may be noted that each mixing operation takes 12 time units for homogeneous
mixing.\hfill$\blacksquare $
\end{example}

\noindent
We introduce a compact symbolic encoding of the actuation sequence generated by a synthesis tool in 
an intermediate language, which is expressive enough to capture all commonly used fluidic operations. The semantics of 
the instruction set is described in Table~\ref{tab:instructions}. \x{The instruction set considered in 
Table~\ref{tab:instructions} is designed to capture the most common operation types, as witnessed in common bioassay 
protocols. For more enhanced protocol instructions, this instruction set can be augmented with additional constructs. In 
Section VII, we have shown an example of an error-tolerant synthesis strategy, which is being studied today in the 
context of cyber-physical systems. We have added two new features (optical detection and conditionals) to enhance our 
repertoire of instructions. The proposed correctness checking method is flexible enough to accommodate such 
additional instructions as required.}

\ctable[
caption = Fluidic instructions and their interpretations,
label = tab:instructions,
pos = !h,
width = 0.48\textwidth,
framesep=0pt,
framerule = 1pt,
doinside = \scriptsize
]{p{70pt}X}{
\tnote{adjacent left, right, top and bottom cells}
}{
\FL
\multicolumn{1}{c}{\textbf{Encoding}} & \multicolumn{1}{c}{\textbf{Architectural description}} \ML
$dim (r,c)$ & Dimension of the biochip, $r$ rows and $c$ columns, each cell is addressed as $(i,j)$, where $1\leq
i\leq r$ and\linebreak $1\leq j\leq c$  \NN
$R(r,c,Name)$ & Reagent reservoir from which droplets can be dispensed. Droplet dispensed from reagent reservoir at
$(r,c)$ and $Name$ denotes reagent's name   \NN
$O(r,c)$/$W(r,c)$ & Output/waste reservoir, where $(r,c)$ denotes the cell from which a droplet can be sent to
output/waste reservoir \ML
\multicolumn{1}{c}{\textbf{Instruction}} &
\multicolumn{1}{c}{\textbf{Fluidic operational description}} \ML
 $d$($r,c$) & Dispense droplet from reservoir at location ($r,c$)   \NN
$m$([$r_1,c_1$]$\to$[$r_2,c_2$]) & Transport droplet at location ($r_1,c_1$) to ($r_2,c_2$). It moves a droplet at
($r_1,c_1$) to one of its 4-neighbors\tmark ($N_4(r_1,c_1)$) i.e.,  $(r_2,c_2) \in N_4(r_1,c_1)$\NN
 $mix$([$r_1,c_1$]$\leftrightarrow$[$r_2,c_2$],\linebreak\hspace*{25pt}$t_{mix},mtype$) & Mix two droplets at
($r_1,c_1$) and ($r_2,c_2$) for $t_{mix}$ time steps. Initial locations are determined by the type of mixer $(mtype)$.
At the end of the mixing operation, two droplets are generated and stored at locations ($r_1,c_1$) and ($r_2,c_2$)
respectively \NN
 $waste(r,c)$/$output(r,c)$ & Dispense droplet at ($r,c$) to its adjacent waste/output reservoir  \NN
 $end$ & End of bioassay \ML
\multicolumn{1}{c}{\textbf{Encoding }} &\multicolumn{1}{c}{\textbf{Bioassay parameter}} \ML
$accuracy$ $n$ & Given $N$ input reagents, each concentration is represented as \linebreak $C=\frac{a_1R_1 +
a_2R_2+\ldots+a_NR_N}{2^n}$, where $\sum_{k=1}^N a_k = 2^n$ and $a_k$ is a positive integer and
$R_k$ is an input reagent for $k= 1,2,\ldots, N$
\LL
}

\noindent
Fig.~\ref{fig:synopsis} shows an example encoding of a synthesized actuation sequence. The first three lines 
define the dimension of the biochip, desired accuracy of concentration, and reservoir locations respectively. Each
instruction line contains one or more symbolic fluidic instruction, separated by space. The instructions occurring on
the same line are concurrent, i.e., they are executed simultaneously. The numeric label prefixing each instruction line 
stands for the start time of the instructions at that line. We assume that every primitive instruction, except 
`$mix([r_1,c_1]\leftrightarrow [r_2,c_2],t_{mix},mtype)$', takes  unit time.

\begin{example}
\normalfont
Interpretations of the different instructions are illustrated in Fig.~\ref{fig:example}, where $S,B,W$ and $O$
 denote the sample dispenser,  buffer dispensers, waste  reservoir and output reservoir, respectively.  The first three
lines of the synthesized output in Fig.~\ref{fig:synopsis} describe the architectural description of the DMF platform on
which the actuation sequences will execute. In this case, the biochip has 5 rows and 4 columns, i.e., it has 20 cells
$(dim\:5\, 4)$. Two reagent reservoirs, namely sample (S) and buffer (B) dispense droplets to $(1,1)$ and $(1,4)$
locations, respectively. Analogously, one waste and one output reservoir dispense droplets from $(5,4)$ and $(5,1)$
respectively. These dispense operations are denoted as $R(1,1,S)\: R(1,4,B)\: O(5,1)\: W(5,4)$. The accuracy level is 
set to 5, i.e., each concentration factor is approximated as $C = \frac{a_1S+a_2B}{2^5}$, where $a_1+a_2=2^5$.

We now describe the execution semantics of the symbolic fluidic operations on the DMFB platform
discussed in the previous paragraph. Some snapshots over different time steps are shown in
Fig.~\ref{fig:example}, where instructions are shown beneath the figure and the time interval is shown on top of the
figure. At $t=1$, two droplets are dispensed from sample (S) and buffer (B) reservoirs. At $t=2$ and $t=3$, both the
droplets are simultaneously moved one cell downward. At $t=4$, a buffer droplet is dispensed to location $(1,4)$ and a 
$1\times 4$ mixer is instantiated, which remains active for the next 12 time steps. At $t=17$, mixing/splitting is 
completed and two droplets are present at locations $(3,1)$ and $(3,4)$. The droplet at $(3,4)$ is then moved 
downward. In the next time step, i.e., $t=18$, droplets at $(1,4)$ and $(4,4)$ are simultaneously moved to $(2,4)$ and 
$(5,4)$ locations respectively. At $t=19$, the droplet at $(2,4)$ is moved to $(3,4)$ and the waste droplet at $(5,4)$ 
is dispatched to the waste reservoir. At a subsequent time $t=20$, a new mixing operation is performed between 
the droplets at  $(3,1)$ and $(3,4)$ using a $1\times 4$ mixer which remains active during the subsequent 12 
time steps. The remaining instructions of the synthesized bioassay (Fig.~\ref{fig:synopsis}) are executed as specified, 
and have similar execution semantics.\hfill $\blacksquare $
\end{example}

\subsection*{Parameters of the target DMF architecture}
\noindent
In order to perform automated correctness checking of the synthesized bioassay, it is essential to know the 
architectural description of the target DMF platform where the synthesized
bioassay will be executed such as its 
dimensions, the reservoir locations, and their contents (reagent, output, waste).

\subsection*{Design constraints}
\noindent
Automated synthesis tools generate actuation sequences that have to satisfy some design constraints. As an example,
both static and dynamic fluidic constraints ~\cite{KrishSu} (Section~\ref{subsec:FC}) need to be
satisfied for unintentional mixing of droplets. Another example of a design constraint is the maximum bioassay
completion time $(T_{max})$, which dictates that the synthesized bioassay should be completed within a specified time
limit. In order to ensure correct realization of the input protocol, there are several other design constraints that
need to be satisfied.

\subsection{Constraint modeling}
\label{subsec:constraint_modeling}
\noindent
We now describe the overall constraint modeling and verification scheme. We create Boolean formulas for verifying each 
of the fluidic instructions. Logical operators are used for negation $(\neg)$, and $(\wedge)$, or $(\vee)$  of Boolean 
variables. Let us consider a two-dimensional biochip of size $r\times c$, where $r$ and $c$ denote the number of rows 
and columns of the target DMFB,  respectively.

We define an
indicator Boolean
variable $x_{i,j}^t$  denoting the state of a biochip cell at location $(i,j)$, where $1\leq i \leq r$, $1\leq j \leq c$
and $t$ is the time index. We encode  a biochip description symbolically at time instant $t$ using $B_{r\times c}^{t}$,
which is the conjunction of
all state variables at time $t$. We define a few Boolean functions and auxiliary tables that represent different
constraints and description of on-chip reservoirs, active mixers in the layout.

\begin{figure}[!h]
\centering
\includegraphics[scale=.65]{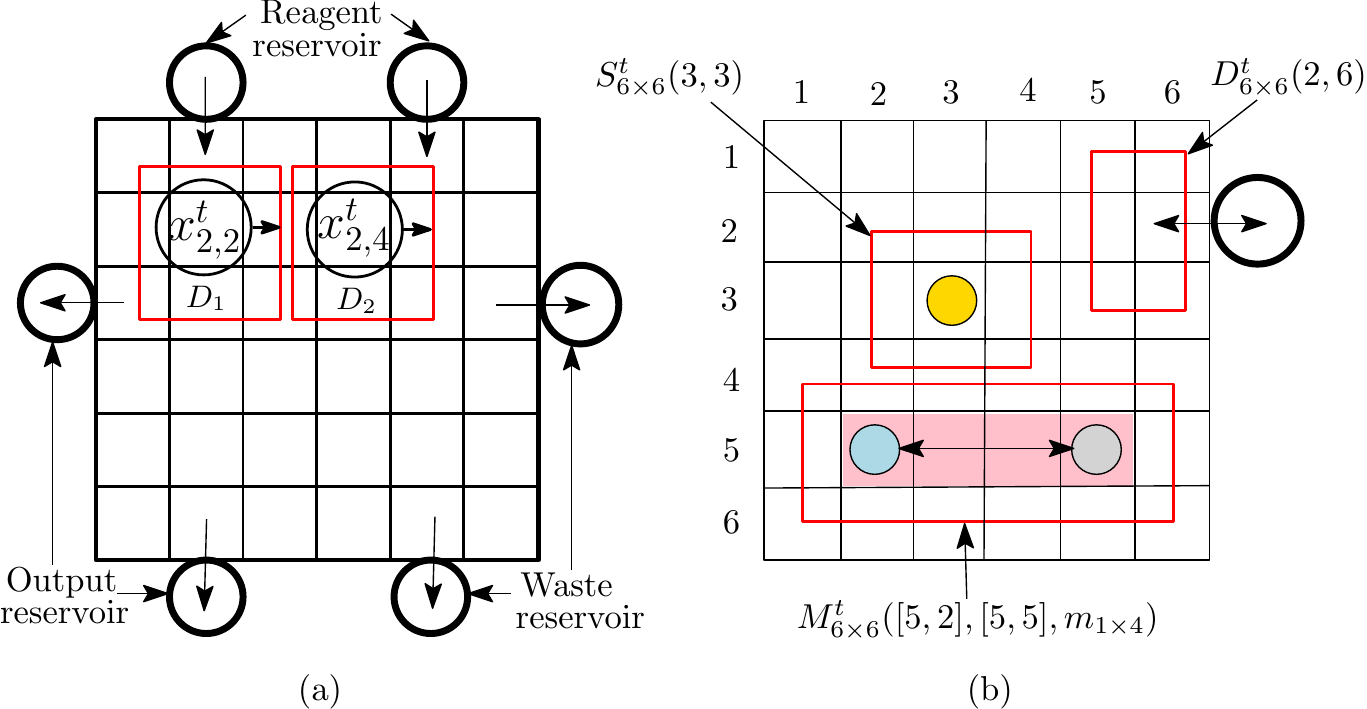}
\caption{(a) Boolean variables for biochip description and fluidic constraints. (b) Notation for design constraints.}
\label{fig:fc_symvar}
\end{figure}

\subsection*{Modeling Fluidic constraints}
\label{subsec:FC}
\noindent
During droplet routing, a minimum spacing between droplets must be maintained to prevent accidental mixing, unless 
droplet merging is desired. Such fluidic constraints (FC)~\cite{KrishSu} are usually classified as static or dynamic, 
depending on the type of operation it relates to.

\begin{definition}{\bf [Static Fluidic constraint:]}
\normalfont
 If a droplet is on cell  $(i, j)$ at time $t$, then no other droplet can reside in the 8-neighboring cells of $(i, j)$ 
at time $t$, \x{denoted as $N_8(i,j)$}\footnote{the 4
diagonal neighbors + 4 vertical and horizontal neighbors as  earlier }.\hfill
$\blacksquare $
\end{definition}

\noindent
For example, in Fig.~\ref{fig:fc_symvar}(a), the static fluidic constraint for the droplet on cell $(2,2)$ at $t$ is
shown as a surrounded box of cells which should not be used by any other droplet at $t$.\linebreak

\noindent
Static fluidic constraint for a droplet on cell $(i,j)$ can be expressed as:
\x{
\begin{align}
\label{eq:sfc}
S_{r\times c}^t(i,j) = x_{i,j}^t &\wedge_{(i_1,j_1)\in N_8(i,j)} \neg x_{i_1,j_1}^t  \\
&\qquad{} \text{where } 1\leq i_1\leq r, 1\leq j_1 \leq c \nonumber
\end{align}
}
\noindent
For example, in Fig.~\ref{fig:fc_symvar}(b), the static fluidic constraint for the droplet at $(3,3)$ at $t$ will be
modeled as  $S_{6\times 6}^t(3,3) = \neg x_{2,2}^t\wedge \neg x_{2,3}^t\wedge \neg x_{2,4}^t\wedge\neg x_{3,2}^t\wedge 
x_{3,3}^t\wedge\neg x_{3,4}^t\wedge\neg x_{4,2}^t\wedge \neg x_{4,3}^t\wedge \neg x_{4,4}^t$\linebreak

\begin{definition}{\bf [Dynamic Fluidic constraint:]}
\normalfont
Every movement operation at an instant $t+1$ must honor the 8-neighborhood isolation of all other droplets at time
$t$.
\hfill$\blacksquare $
\end{definition}

\noindent
In other words, the activated cell for droplet $D_i$ cannot be adjacent to $D_j$, otherwise, there is more than one
activated neighboring cell for $D_j$, which may lead to errant fluidic operations. Consider Fig.~\ref{fig:fc_symvar}(a)
with two droplets $D_1$ and $D_2$  at locations $(2,2)$ and $(2,4)$ respectively. If both $D_1$ and $D_2$ want to move 
one cell right at $t+1$, two neighboring cells of $D_2$  are actuated simultaneously. Although the static fluidic 
constraint  is satisfied, this situation may still lead to unintentional mixing of droplets at location $(2,3)$. As a 
result, $D_1$ moves rapidly to $(2,3)$ while the droplet $D_2$ moves slowly towards $(2,5)$.

Modeling of dynamic fluidic constraints is non-trivial since it depends on the exact type of the movement operation.
We describe below two other types of constraints that relate to dispensing and mixing.

\begin{definition}{\bf [Dispensing constraint:]}
\normalfont
Before dispensing a droplet at $(i,j)$,  no droplet should be present either at $(i,j)$ or in its 8-neighborhood. 
\hfill$\blacksquare $
\end{definition}
\noindent
This constraint can be expressed as:
\x{
\begin{align}
\label{eq:dispense}
D_{r\times c}^t(i,j)= \neg x_{i,j}^t &\wedge_{(i_1,j_1)\in N_8(i,j)} \neg x_{i_1,j_1}^t\\
&\qquad{} \text{where } 1\leq i_1\leq r, 1\leq j_1\leq c \nonumber
\end{align}
}
\noindent
In Fig.~\ref{fig:fc_symvar}(b),  $D_{r\times c}^t(2,6)$ denotes the constraint that needs to be satisfied before
dispensing a droplet from a reservoir to cell $(2,6)$ at time $t$. Hence, $D_{r\times c}^t(2,6)= \wedge_{(1\leq
i\leq 3)}\neg x_{i,5}^t\wedge\neg x_{i,6}^t$.

\begin{definition}{\bf [Mixer constraint:]}
\normalfont
No droplet should be present in the adjacent cells of active mixers.\hfill$\blacksquare $
\end{definition}
\noindent
This constraint is needed to prevent  unintentional droplet mixing. The expression for this constraint
varies with the type of the mixer ($mtype$). Intuitively, this  captures the restriction that the droplets
to be mixed should be available at the mixer-input specified locations, while also honoring the neighborhood
constraint in the adjacency of the mixing locations.

Let $M_{r\times c}^t([r_1,c_1],[r_2,c_2],mtype)$ denote the mixer constraint for an active mixer. Note that for 
each type of mixer, it takes two input droplets at location $(r_1,c_1)$ and $(r_2,c_2)$ and produces 
two output droplets at the same locations. \x{For simplicity,  we have considered only $1\times 4$ ($mtype = 14$) and 
$4\times 1$ ($mtype = 41$) mixers.
\begin{example}
\normalfont
For a $1\times 4$ linear array, i.e.,  $mtype=14$,  the mixer constraint can be expressed as:
\begin{equation}
\label{eq:mixer14}
M_{r\times c}^t([r_1,c_1],[r_2,c_2],14) = S_{r\times c}^t(r_1,c_1)\wedge S_{r\times c}^t(r_2,c_2)
\end{equation}
In Fig.~\ref{fig:fc_symvar}(b), the mixer constraint $M_{6\times 6}^t([5,2],[5,5],14) $ for a $1\times 4$
mixer with two droplets at locations $(5,2)$ and $(5,5)$ can be expressed as  $\wedge_{(1\leq j\leq 6)}\neg
x_{4,j}^t\wedge\neg x_{5,1}^t\wedge x_{5,2}^t\wedge\neg x_{5,3}^t\wedge\neg x_{5,4}^t\wedge x_{5,5}^t\wedge\neg
x_{5,6}^t\wedge_{(1\leq j\leq 6)}\neg x_{6,j}^t$.\hfill $\blacksquare$
\end{example}
}
\x{
In the special case of a $1 \times 4$ or $4 \times 1$ mixer, where the two input/output droplets are in the leftmost 
and rightmost cells, the mixer constraint can be described by the conjunction of the static fluidic constraints of the 
two droplet locations. However, for other types of mixers, the constraint in Eqn.~\ref{eq:mixer14} needs to be suitably 
modified for preventing any interference with other droplets. }

\noindent
A table $T_{reservoir}$ stores the on-chip reservoir descriptions as $(i,j,rtype)$, where $(i,j)$ is the dispenser
location and $rtype$ is the reservoir type (reagent, output, waste). \x{For correctness checking 
of each dispense operation, $T_{reservoir}$ is accessed for verifying the correct dispense location.} Similarly, a 
table $T_{mixer}$ maintains the descriptions of active mixers in the form of  $([r_1,c_1],[r_2,c_2],t_s,t_e,mtype)$, 
where $[r_1,c_1],[r_2,c_2]$ are the locations of the two droplets to be mixed with an $mtype$ mixer and $t_s,t_e$ are 
start and end times of a mixing operation. \x{In our incremental correctness checking procedure, we change the DMF 
state depending on the current state and the set of fluidic operations to be executed at the subsequent time 
instants. As mixer instantiation spans over multiple time steps, we need to store active mixers in $T_{mixer}$ so that 
the necessary updating of DMF states can be made on completion of an active mixing at a particular time step. } 
Table~\ref{tab:constraints} summarizes the notations we are using.

\ctable[
caption = Notation,
label = tab:constraints,
pos = !h,
width = 0.48\textwidth,
framesep=0pt,
framerule = 1pt,
doinside = \scriptsize
]{p{25pt}X}{
}{
\FL
\multicolumn{1}{c}{\textbf{Boolean function}} &\multicolumn{1}{c}{\textbf{Constraint}} \ML
$B_{r\times c}^{t}$ & Symbolic description of biochip of size $r\times c$ at time $t$.\NN
$S_{r\times c}^t(i,j)$ & Defined on state variables at time $t$, that represents the static FC of a droplet at $(i,j)$
in a biochip of size $r\times c$\NN
$D_{r\times c}^t(i,j)$ & Represents FCs that must be satisfied before dispensing droplet at $(i,j)$\NN
$M_{r\times c}^t([r_1,c_1],$\linebreak $ [r_2,c_2], mtype)$ & Represents static FC for an active mixer of type $mtype$
at time $t$\ML
\multicolumn{1}{c}{\textbf{Table}} & \multicolumn{1}{c}{\textbf{Description}} \ML
$T_{mixer}$ & Maintains active mixers. Each entry is in the form of  $([r_1,c_1],[r_2,c_2],t_s,t_e,mtype)$, where
$[r_1,c_1],[r_2,c_2]$ are the positions of the two droplets to be mixed with a $mtype$ mixer and $t_s,t_e$ are start and
end times of the mixing operation\NN
$T_{reservoir}$ & Stores the description of on-chip reservoir. Each entry is of the form $(i,j,rtype)$, where $(i,j)$ is
the dispense location and $rtype$ is the reservoir type (reagent, output, waste)
\LL
}

\subsection*{Modeling the initial configuration}
\noindent
The first three lines of the synthesized output describe the architectural description of the DMF
platform on which the input bioassay is to be executed. \x{From the architectural description of the synthesized 
output, on-chip reservoir descriptions are stored in $T_{resrevoir}$ along with its type (input/output/waste 
reservoir). Moreover, the mixer table is initialized, i.e., $T_{mixer} = \phi$, as the biochip has no droplets to be 
mixed initially $(t=0)$. Hence, the symbolic representation of the  biochip at $t=0$ is
\begin{equation}
\label{eq:init}
B_{r\times c}^{0}=\wedge_{(1\leq i\leq r, 1\leq j\leq c)}\neg x_{i,j}^0 
\end{equation}
}
\noindent
In subsequent time steps, some fluidic operations are executed, which lead to a change in the description. In general, 
we  have $B_{r\times c}^{t}$ and some fluidic operation executed at $t+1$. We need to verify whether these operations  
are  valid on the current configuration ($B_{r\times c}^{t}$) and change the biochip to an allowable configuration at 
$t+1$, i.e., $B_{r\times c}^{t+1}$ is valid at $t+1$.

\subsection*{Modeling of fluidic operations}
\label{sec:mix}
\noindent
We now illustrate the modeling of each fluidic operation executed at $t+1$ on $B_{r\times c}^t$. In
the illustrative examples, we only list the \textit{true} variables explicitly in the symbolic representation of the
biochip. All other negated variables are assumed to be implicitly present.\linebreak

\noindent
\textit{Dispense of a reagent droplet -- $d(i,j)$:} Before dispensing a droplet from a reagent reservoir to a cell
$(i,j)$, one has to ensure that the droplet is being dispensed from a  reagent reservoir. This can be checked by a 
look-up on $T_{reservoir}$. Also, after dispensing the droplet at $(i,j)$, no fluidic violation should occur. This can 
be verified using the expression:
\begin{equation}
\label{eq:dispense}
E_d^{FC} = D_{r\times c}^t(i,j) \wedge B_{r\times c}^{t}
\end{equation}
\noindent
If $E_d^{FC}$ is $true$, the  fluidic constraints are satisfied. The truth value checking procedure has been elaborated
in Section~\ref{sec:implementation}. It can be easily verified that  $E_d^{FC}$ is $true$ only when there is no droplet
present at the location $(i,j)$ and  its adjacent locations at time $t$. It is important to ensure that only one
droplet is dispensed on a particular location at $t+1$. This can be done in the following way.

After dispensing a droplet to a valid location at a time step $t+1$, it is important to check that no droplet is 
dispensed on the same location at the same time step. For ensuring this, we need to modify $B_{r\times c}^t$ to reflect 
that the droplet is on $(i,j)$ at $t$. In doing so, we need to update $B_{r\times c}^{t}$ with $x_{i,j}^{t}$. 
After this modification, if we want to dispense another droplet at $(i,j)$, it violates the fluidic constraints since 
the updated $B_{r\times c}^t$ shows a droplet  present at $(i,j)$ at time $t$. Finally, $B_{r\times c}^{t+1}$ is updated 
with $x_{i,j}^{t+1}$ for representing the dispensed droplet on the biochip at $t+1$. We explain this on the 
following example.

\begin{figure}[!h]
\centering
\includegraphics[scale=.75]{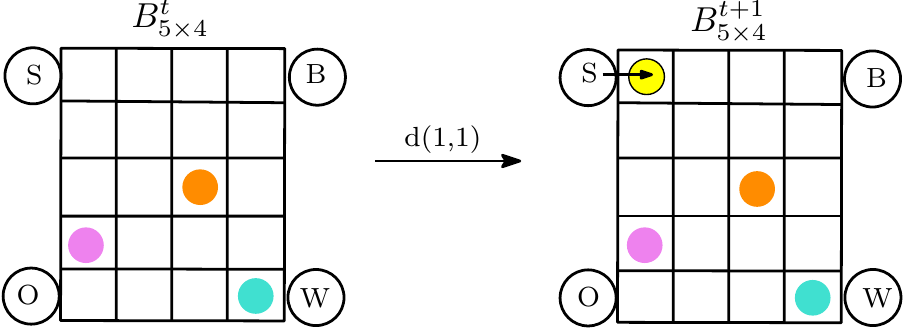}
 \caption{Verifying the dispense operation}
\label{fig:dispense}
\end{figure}

\begin{example}
\normalfont
In Fig.~\ref{fig:dispense}, the biochip description at $t$ is $B_{5\times 4}^t = x_{3,3}^t \wedge x_{4,1}^t \wedge
x_{5,4}^t$ (negated variables are implicit) and at $t+1$, a dispense instruction $d(1,1)$ is to be executed. $E_d^{FC}$ 
is $true$, i.e, fluidic constraints are preserved. We must update $B_{5\times 4}^t$ with $x_{1,1}^t$ to prevent multiple
dispense on the same location $(1,1)$ at $t+1$. Hence, $B_{5\times 4}^t$ becomes $x_{1,1}^t \wedge x_{3,3}^t \wedge
x_{4,1}^t \wedge x_{5,4}^t$ i.e., $\neg x_{1,1}^t$ is removed and  $x_{1,1}^t$ is added. \x{Note that after updating 
$B_{5\times 4}^t$, any operation trying to dispense a droplet on the same location is prohibited.} Now, 
consider a case where a droplet exists on $(2,2)$ at $t$. $B_{5\times 4}^t$ will be $x_{2,2}^t \wedge x_{3,3}^t \wedge 
x_{4,1}^t \wedge x_{5,4}^t$ and $ D_{5\times 4}^t(1,1)$ is $\neg x_{1,1}^t \wedge \neg x_{1,2}^t \wedge \neg x_{2,1}^t 
\wedge \neg x_{2,2}^t$. Note that $B_{5\times 4}^t\wedge D_{5\times 4}^t(1,1)$ is $false$ and therefore, generates a 
conflict. Hence, this dispense is not safe. $B_{5\times 4}^{t+1}$ becomes
$x_{1,1}^{t+1} \wedge x_{3,3}^{t+1} \wedge x_{4,1}^{t+1} \wedge x_{5,4}^{t+1}$. \hfill $\blacksquare$
\end{example}

\noindent
\textit{Dispensing a droplet to a waste reservoir --  $waste(i,j)$:}
This is similar to \textit{d($i,j$)}, except that waste reservoirs are considered instead of reagent reservoirs. This
can be expressed as:
\x{
\begin{equation}
\label{eq:waste}
E_{waste}^{FC}=x_{i,j}^t \wedge B_{r\times c}^{t}
\end{equation}
}
\noindent
$E_{waste}^{FC}$ is $true$ if a droplet is present on $(i,j)$ before dispensing. \x{As earlier, in this case as well, we 
must update $B_{r\times c}^t$ with $\neg x_{i,j}^t$ to prevent multiple dispense to the waste reservoir from the same 
location $(i,j)$ at $t+1$.} $B_{r\times c}^{t+1}$ is updated with $\neg x_{i,j}^{t+1}$ to capture the fact that the cell 
is now available, since the droplet occupying the cell has been dispensed to a waste reservoir.\linebreak

\noindent
\textit{Dispensing a droplet to an output reservoir -- $output(i,j)$:}
The formulation for this is similar to  \textit{waste($i,j$)}, except that  output reservoirs are considered. This can 
be expressed as:
\x{
\begin{equation}
\label{eq:output}
E_{output}^{FC}  = x_{i,j}^t \wedge B_{r\times c}^{t}
\end{equation}
}
\noindent
As earlier, $E_{output}^{FC}$  must be $true$ before dispensing. \x{Analogously, $B_{r\times c}^t$  
and $B_{r\times c}^{t+1}$ are updated with $\neg x_{i,j}^{t}$ and $\neg x_{i,j}^{t+1}$ respectively.}\linebreak

\noindent
\textit{Transportation of a droplet -- $m( [r_1,c_1]\to[r_2,c_2] )$:}
\x{
The move instruction $m$ transports a droplet from a valid DMFB location $(r_1,c_1)$ to one of its four neighbors 
$(r_2,c_2)$ i.e., $(r_2,c_2) \in N_4(r_1,c_1)$. The static and dynamic fluidic constraints need to be
satisfied in the move operation. The fluidic constraint for this operation can be expressed as:
\begin{equation}
\label{eq:move}
E_m^{FC}= x_{r_1,c_1}^t \wedge E\wedge B_{r\times c}^{t}
\end{equation}
\noindent
The expression  $E$ is used for checking dynamic fluidic constraints and is defined as follows:
\begin{equation*}
\scriptsize
E =
\begin{cases}
\neg x_{r_2-1,c_2+1}^t\wedge \neg x_{r_2,c_2+1}^t\wedge \neg x_{r_2+1,c_2+1}^t & 
\mbox{for right ($\rightarrow$) move } \\
\neg x_{r_2-1,c_2-1}^t\wedge \neg x_{r_2,c_2-1}^t\wedge \neg x_{r_2+1,c_2-1}^t & 
\mbox{for left ($\leftarrow$) move } \\
\neg x_{r_2+1,c_2-1}^t\wedge \neg x_{r_2+1,c_2}^t\wedge \neg x_{r_2+1,c_2+1}^t & 
\mbox{for bottom ($\downarrow$) move } \\
\neg x_{r_2-1,c_2-1}^t\wedge \neg x_{r_2-1,c_2}^t\wedge \neg x_{r_2-1,c_2+1}^t & 
\mbox{for top ($\uparrow$) move } \\
\end{cases}
\end{equation*}
\noindent
Note that $x_{r_1,c_1}^t \wedge B_{r\times c}^{t}$ is $true$ when a droplet is present on $(r_1,c_1)$ at time $t$.
Analogously, the dynamic fluidic constraint is satisfied if $E\wedge B_{r\times c}^{t}$ is $true$ i.e., no droplet is
present at the neighboring locations of the move destination. $B_{r\times c}^{t+1}$ is updated with $\neg
x_{r_1,c_1}^{t+1}\wedge x_{r_2,c_2}^{t+1}$ for reflecting the effect of the move operation at $t+1$. It may be noted 
that the neighborhood constraint for $(r_2, c_2)$ is already ensured partially by the fact that the present 
configuration with a droplet at $(r_1, c_1)$ at time $t$ satisfies all static fluidic constraints.}

\begin{figure}[!h]
\centering
\includegraphics[scale=.75]{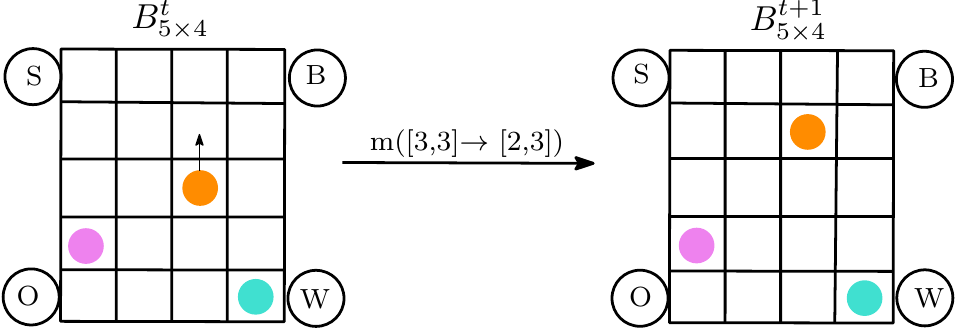}
  \caption{Verifying the move operation.}
  \label{fig:move}
\end{figure}

\begin{example}
\normalfont
Consider the example shown in Fig.~\ref{fig:move} where $B_{5\times 4}^t = x_{3,3}^t \wedge x_{4,1}^t
\wedge x_{5,4}^t$. At $t+1$, a  move instruction $m([3,3]\rightarrow [2,3])$ is executed. Note that the move
destination is one of the 4-neighbors of $(3,3)$. \x{$E_{m}^{FC} =  x_{3,3}^t\wedge  (\neg x_{1,2}^t\wedge \neg 
x_{1,3}^t\wedge \neg x_{1,4}^t)\wedge (x_{3,3}^t \wedge x_{4,1}^t \wedge x_{5,4}^t)$ is $true$.} So, a droplet is 
present on $(3,3)$ at $t$ and when it is moved to $(2,3)$ no fluidic constraints are violated. 

Consider the case where $m([3,3]\rightarrow [3,2])$ is executed at $t+1$ instead of $m([3,3]\rightarrow [2,3])$. \x{Note
that  $E_{m}^{FC} =  x_{3,3}^t \wedge  (\neg x_{2,1}^t\wedge \neg x_{3,1}^t\wedge \neg x_{4,1}^t)\wedge (x_{3,3}^t 
\wedge x_{4,1}^t \wedge x_{5,4}^t)$ is $false$ because it conflicts on variable $x_{4,1}^t$.} Hence, the instruction 
$m([3,3]\rightarrow [3,2])$ violates the fluidic constraint. \hfill$\blacksquare $
\end{example}

\noindent
\textit{Mixing operation -- $mix([r_1,c_1]\leftrightarrow[r_2,c_2],t_{mix},mtype$):}
\noindent
The  expression for a mixer depends on the mixer type $(mtype)$. Depending on the type, the verification of
the mixer instantiation at $t+1$ should ensure that two droplets must be present at the correct locations ($r_1,c_1)$ 
and $(r_2,c_2)$ at $t$. Additionally, $((r_1,c_1),(r_2,c_2),t+1,t+1+t_{mix},mtype)$ should be inserted in the active 
mixer table $T_{mixer}$. Moreover, $B_{r\times c}^{t+1}$ has to be updated for indicating the instantiation of the 
mixer. It is important to note that when mixing ends at $t+t_{mix}+1$, $B_{r\times c}^{t+t_{mix}+1}$ should reflect 
that mixing is over and two new droplets  are created at $(r_1,c_1)$ and $(r_2,c_2)$.

\begin{example}
\normalfont
In Fig.~\ref{fig:fc_symvar}(b), a $1\times 4$ mixer is instantiated with two droplets at locations $(5,2)$
and $(5,5)$. In case of a $1\times 4$ mixer, the presence of two droplets at locations $(5,2)$ and $(5,5)$
can be easily ensured by checking the truth value of \x{$M_{6\times 6}^t([5,2],[5,5],14)$,} which is $S_{6\times
6}^t(5,2) \wedge S_{6\times 6}^t(5,5) \wedge B_{6\times 6}^t$. The  expression \x{$M_{6\times 6}^t([5,2],[5,5],14)$} 
remains $true$ throughout $t+1$ to $t+t_{mix}+1$.\hfill $\blacksquare $
\end{example}

\section{Verifying protocols on pin-constrained DMFB}
\label{sec:pin_constrained}
 \begin{figure*}[th]
\centering
\includegraphics[scale = .8]{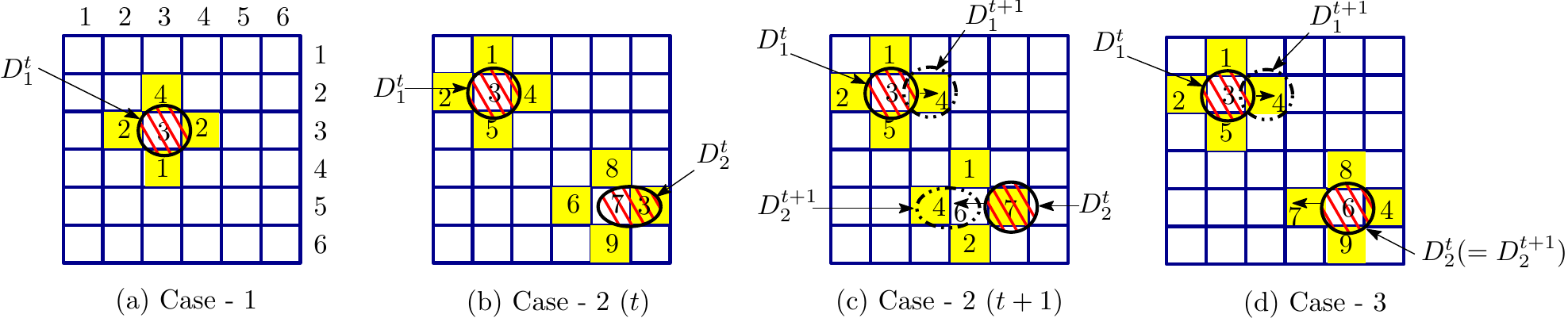}
\caption{Design rules for pin-constrained DMFB: (a) Case-1 (b) Case-2 (at time $t$) (c) Case-2 (at time $t+1$) (d)
Case-3}
\label{fig:pin_example}
\end{figure*}

\noindent
A general-purpose reconfigurable DMFB has a dedicated pin for controlling the actuations of each individual
electrode. Hence, one can navigate each droplet to and from any electrode independently without affecting any other
droplet sitting or moving elsewhere.  As a trade-off between the flexibility of droplet 
movement and pin count, pin-constrained DMFBs~\cite{XuPinConstrained,LuoPinConstrained} may be
deployed to reduce the number of control pins (single pin controls a group of electrodes) without affecting desired
droplet movement.

Although earlier design methodologies~\cite{XuPinConstrained,LuoPinConstrained} can reduce the pin count significantly, 
they enforce several constraints on droplet movement. In this section, we describe the verification procedure for 
ensuring correct movement of two droplets on a pin-constrained DMFB. Moreover, for more than two droplets, one needs to 
consider the interference among them. 

\subsection{Design rules for pin-constrained DMFB}
\noindent
The pin assignment problem satisfying the constraints for simultaneous movement of two or more droplets was studied 
earlier~\cite{XuPinConstrained}. Our proposed verification scheme ensures that the fluidic instructions to be executed 
at time step $t$ satisfy all such constraints. Assume that $Pin(L)$ denotes the set of pins assigned to the electrodes 
of the set $L$ of cells, where $L\subseteq \{(x,y): 1\leq x \leq r, 1\leq y \leq c\}$, in a $(r\times c)$ DMFB. Let 
$N_4(x,y)$ denote the set of directly adjacent four neighbors of location $(x,y)$ as earlier. Let us consider now the 
design constraints on a case-by-case  basis, assuming at $t$, the locations of two droplets are $D_1^t=(x_1^t,y_1^t)$ 
and $D_2^t=(x_2^t,y_2^t)$ respectively.  At $t+1$, both droplets move by one cell and their locations become 
$D_1^{t+1}=(x_1^{t+1},y_1^{t+1})$ and $D_2^{t+1}=(x_2^{t+1},y_2^{t+1})$ respectively.

\begin{case}
\label{case:1}
\normalfont
At any time instant, distinct pins are required to control the four neighborhood electrodes of any droplet.\hfill
$\blacksquare$
\end{case}
\noindent
\x{Let a droplet be placed on location $(x^t,y^t)$ at time $t$. The above constraint can be modeled as follows.
\begin{equation}
\scriptsize
\label{eq:case1}
Pin(\{(x_1^t,y_1^t)\}) \bigcap_{\substack{(x_1^t,y_1^t),(x_2^t,y_2^t)\in N_4(x^t,y^t)\\ 
(x_1^t,y_1^t)\neq(x_2^t,y_2^t)}} Pin(\{(x_2^t,y_2^t)\}) = \phi
\end{equation}
}
This is needed to ensure  possible movement of a droplet into its 4-neighborhood. This can be checked by
analyzing the pairwise intersection of the electrode controlling pins of the 4-neighborhood of each droplet.
\begin{example}
\normalfont
In Fig.~\ref{fig:pin_example}(a), at time $t$, a droplet is placed on $(3,3)$. Note that $Pin(\{(3,3)\}) = 3$,
$N_4(3,3) = \{(2,3),(3,2),(4,3),(3,4)\}$ and $Pin(\{(3,2)\}) = Pin(\{(3,4)\}) = 2$. Hence, two opposite neighborhood
electrodes of the droplet at $(3,3)$ share the same pin. This sharing will not cause any violation as long as pin 2 is
inactive. In order to enable the left/right movement operation of the droplet at $(3,3)$, we need to activate pin 2
 and deactivate pin 3  simultaneously.  As a result, the droplet at $(3,3)$ will be split into two droplets
unintentionally. Moreover, if two diagonal electrodes share the same pin, the droplet will be moved (instead of
splitting) in between two diagonal electrodes that are controlled by the same pin. \hfill $\blacksquare$
\end{example}

\begin{case}
\label{case:2}
\normalfont
At any time instant, the pin controlling the electrode of any droplet cannot be shared with any pins that
control the 4-neighborhood electrodes of any other droplet.\hfill $\blacksquare$
\end{case}
\noindent
In the case of $D_1^t$ and $D_2^t$, we can model the above constraint as:
\begin{equation}
(a):\:Pin(\{(x_1^t,y_1^t)\}) \cap Pin(N_4(x_2^t,y_2^t)) = \phi
\end{equation}
Interchanging the role of $D_1$ and $D_2$, we get
\begin{equation}
(b):\:Pin(N_4(x_1^t,y_1^t)) \cap Pin(\{(x_2^t,y_2^t)\}) = \phi
\end{equation}

\begin{example}
\normalfont
In Fig.~\ref{fig:pin_example}(b), two droplets are placed on $(2,2)$ and $(5,5)$ respectively, at time $t$.
Note that $Pin(\{(2,2)\}) = 3$, $N_4(5,5) = \{(4,5),(5,4),(6,5),(5,6)\}$, $Pin(N_4(5,5))
=\{8,6,9,3\}$. Hence, $Pin(\{(2,2)\})\cap Pin(N_4(5,5)) = \{3\}$ i.e., at time $t$ one of the 4-neighbors of droplet
$D_2^t$ shares the same pin $(3)$ of droplet $D_1^t$. Since both pins 3 and 7 are actuated with a high signal, $D_2^t$ 
may be unintentionally stretched and stay between electrodes $(5,5)$ and $(5,6)$.\hfill $\blacksquare$
\end{example}

\noindent
Similarly at $t+1$, Case~\ref{case:2} can be encoded as:
\begin{equation}
(c):\:Pin(\{(x_1^{t+1},y_1^{t+1})\}) \cap Pin(N_4(x_2^{t+1},y_2^{t+1})) = \phi
\end{equation}
Interchanging the role of $D_1$ and $D_2$, we get
\begin{equation}
(d):\:Pin(N(x_1^{t+1},y_1^{t+1})) \cap Pin(\{(x_2^{t+1},y_2^{t+1})\}) = \phi
\end{equation}

\begin{example}
\normalfont
Let us consider at time $t$, two droplets are placed on $(2,2)$ and $(5,5)$ and at $t+1$, they move to $(2,3)$
and $(5,4)$ respectively (Fig.~\ref{fig:pin_example}(c)). Note that $Pin(\{(2,3)\}) = 4$, $N_4(5,4) =
\{(4,4),(5,3),(6,4),(5,5)\}$, $Pin(N_4(5,4)) = \{1,4,2,7\}$. Hence, $Pin(\{(2,3)\})\cap Pin(N_4(5,4)) = \{4\}$, i.e., at
time $t+1$, one of the 4-neighbors of droplet $D_2^{t+1}$ shares the same pin $(4)$ of droplet $D_1^{t+1}$. When both 
pin 4 and 6 are actuated with a high signal, $D_2^{t+1}$ may be
unintentionally stretched and stay between the electrodes $(5,3)$ and $(5,4)$.\hfill $\blacksquare$
\end{example}

\begin{case}
\normalfont
In order to move two droplets simultaneously at time instant $t+1$, the controlling pins of the destination electrodes, 
cannot be shared with any pins that control the 4-neighborhood electrodes
of the other droplet at time $t$, except when the new locations of the two droplets share the same pin.\hfill
$\blacksquare$
\end{case}

\noindent
Hence, the movement constraint for $D_1^{t+1}$ and $D_2^{t+1}$ can be modeled as:
\begin{equation}
\footnotesize
(a):\:Pin(\{(x_1^{t+1},y_1^{t+1})\}) \cap Pin(N_4(x_2^t,y_2^t)\setminus \{(x_2^{t+1},y_2^{t+1})\})
= \phi
\end{equation}

\noindent
Interchanging the roles of $D_1$ and $D_2$, we get the following constraint that needs to be honored as well:
\begin{equation}
\footnotesize
(b):\:Pin(N_4(x_1^t,y_1^t)\setminus \{(x_1^{t+1},y_1^{t+1})\}) \cap Pin(\{(x_2^{t+1},y_2^{t+1})\})
= \phi
\end{equation}

\begin{example}
\normalfont
Assume, at time $t$, two droplets are placed on $(2,2)$ and $(5,5)$ and in the next time step i.e., at $t+1$, they move
to $(2,3)$ and $(5,4)$ respectively. As shown in Fig.~\ref{fig:pin_example}(d), $Pin(\{(2,3)\}) = 4, N_4(5,5) =
\{(4,5),(5,4),(6,5),(5,6)\}$, $Pin(N_4(5,5)\setminus \{(5,4)\}) = \{8,9,4\}$. Hence, $Pin(\{(2,3)\})\cap
Pin(N_4(5,5)\setminus \{(5,4)\}) = \{4\}$. Note that, at $t+1$, when $D_2^t$ tries to move from $(5,5)$ to $(5,4)$, pin
7, pin 6 and pin 4 are actuated with high, low and high signals respectively. As a consequence, $D_2^t$ is pulled with
nearly equal forces from two adjacent high electrodes (controlled by pin 7 and pin 4) and cannot move
as anticipated. However, at time $t+1$, if $D_2^t$ is moved to $(5,6)$ instead of $(5,4)$,
then two droplets can be moved to their desired locations.\hfill $\blacksquare$
\end{example}

\noindent
So far we have discussed the design constraints that need to be satisfied when two droplets move simultaneously.
Additionally, we need to consider the design constraints when only one droplet moves and other remains at its
location. However, this is a special case of simultaneous movement of two droplets. Without loss of generality, let
us consider that the droplet $D_2^t$ remains in the same position at $t+1$, i.e., $D_2^{t+1}=D_2^t$. If we substitute
$(x_2^t,y_2^t)$ by $(x_2^{t+1},y_2^{t+1})$ in the  cases 2 and 3, we get six additional constraints that need
to be satisfied when only one droplet moves and another remains static.

\subsection{Verifying  fluidic operations on a pin-constrained DMFB}
\noindent
In this subsection, we describe the verification method  for each fluidic instruction executed on a pin-constrained
DMFB. We assume that the fluidic instructions executed at time $t$ do not violate any design constraints.

\subsection*{Dispense a reagent droplet -- $d(i,j)$:}
\noindent
Before dispensing a reagent droplet, we need to ensure that it should not create any interference
with other droplets present at $t$. This constraint can be verified by checking the intersection between the pin
assigned to the electrode of the dispensed location $(i,j)$ and the pin assigned to the neighboring electrodes of each 
droplet. Moreover, the pin assigned to $(i,j)$ should not share any pin with its own neighbors. Otherwise,
undesirable droplet stretching may occur.

Let there be $k$ droplets  $\{(x_1^t,y_1^t),(x_2^t,y_2^t),\ldots,(x_k^t,y_k^t)\}$  present on a DMFB at time
$t$. At $t+1$, a droplet is dispensed from a reagent reservoir to location $(i,j)$. If $Pin(\{(i,j)\})\cap
Pin(N_4(x_l^t,y_l^t)) = \phi$, for $l=1,2,\ldots ,k$ and $Pin(\{(i,j)\})\cap Pin(N_4(i,j)) = \phi$,  it is safe to
dispense a droplet on $(i,j)$ at time $t+1$.

\begin{figure}[!h]
\centering
\includegraphics[scale=.9]{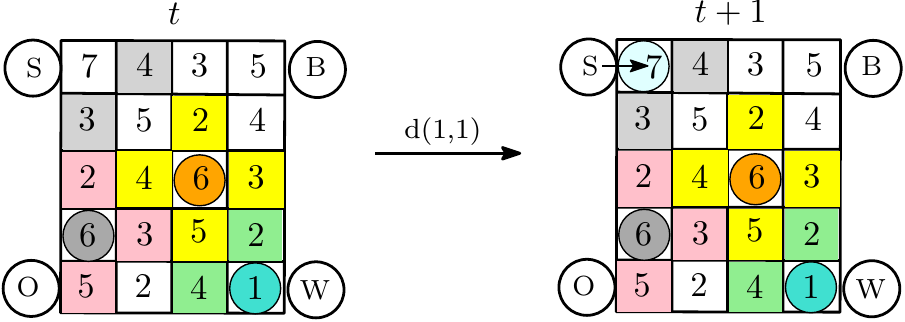}
\caption{Verifying the dispense operation}
\label{fig:dispense_pc}
\end{figure}
\begin{example}
\normalfont
Assume at time $t$, three droplets are present at locations $(3,3),(4,1)$ and $(5,4)$ (Fig.~\ref{fig:dispense_pc}). It
is easy to check that at time $t$, there are no violations, and a droplet can be dispensed on $(1,1)$ at time $t+1$.   
$Pin(N_4(3,3)) = \{2,4,5,3\}$, $Pin(N_4(4,1)) = \{2,5,3\}$, $Pin(N_4(5,4))=\{2,4\}$ and $Pin(N_4(1,1))=\{3,4\}$. Hence, 
the pin  assigned to $(1,1)$ is disjoint with the set of pins assigned to its own neighbors and the neighbors of other 
droplets, i.e., $\{7\} \cap \{2,4,5,3\} = \phi$,  $\{7\}\cap\{2,5,3\} = \phi$, $\{7\}\cap\{2,4\}=\phi$, $\{7\}\cap 
\{3,4\}=\phi$. Hence, the reagent droplet can be safely dispensed on $(1,1)$ at $t+1$.

If at $t+1$, a droplet is dispensed to $(1,4)$ instead of $(1,1)$, we have a violation. Note that, the
pin assigned to $(1,4)$ has a non-null intersection with $Pin(N_4(3,3))$ and $Pin(N_4(4,1))$. Hence, droplet
stretching may occur.\hfill $\blacksquare$
\end{example}

\subsection*{Transporting a droplet to a waste reservoir --  $waste(i,j)$:}
\noindent
This case is straightforward. We need to ensure that the waste droplet to be transported at $t+1$ must not share any pin 
with other droplets present at time $t$. This fact is true because at $t$, no design constraint was violated.

\subsection*{Transporting a droplet to an output reservoir -- $output(i,j)$:}
\noindent
This is similar to $waste(i,j)$.

\subsection*{Transportation of a droplet -- $m( [r_1,c_1]\to[r_2,c_2])$:}
\noindent
The correctness checking of concurrent droplet movements can be decomposed into
checking for all possible pairs of droplet movement. In fact, the verification of simultaneous movement of
a pair of  droplets on a pin-constrained DMFB is equivalent to checking Cases 1-3. Note that we have to check
Case-2 only for time instant $t+1$, because at time $t$, all constraints are satisfied. Moreover, we can suitably
modify Cases 2-3 for verifying the instance when only one droplet is moved and the other remains static. The next 
example illustrates this scenario.

\begin{figure}[!h]
\centering
\includegraphics[scale=.9]{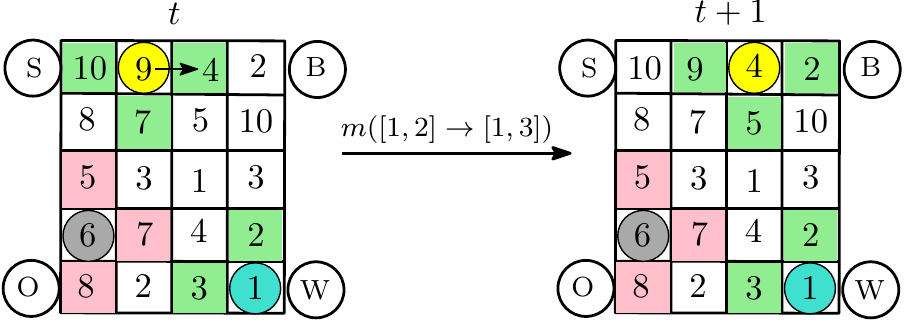}
\caption{Verifying move operation on pin-constrained DMFB}
\label{fig:move_pc}
\end{figure}
\begin{example}
\normalfont
Suppose, at time $t$, three droplets are present on locations $(1,2),(4,1)$ and $(5,4)$ respectively. At $t+1$,
$m([1,2]\rightarrow [1,3])$ will be executed, i.e., droplet at $(1,2)$ will be moved to its right neighbor at $(1,3)$
and other droplets will not move. Note that $Pin(\{(1,2)\})=\{9\}, Pin(N_4(1,2))=\{10,7,4\}, Pin(N_4(1,3))=\{9,5,2\}$ 
and $Pin(\{(4,1)\})=\{6\}, Pin(N_4(4,1))=\{5,8,7\}$. It can be easily verified that the constraint in Case 1 is 
satisfied. The following cases (Case 2 and Case 3) must be satisfied for checking interference of moving droplet with 
the static
droplet on $(4,1)$.

{\footnotesize
\noindent
\textbf{Cases:}
\begin{itemize}
 \item[2(c):] $Pin(\{(1,3\})\cap Pin(N_4(4,1)) = \{4\}\cap\{5,8,7\}=\phi$
 \item[2(d):] $Pin(N_4(1,3))\cap Pin(\{(4,1)\}) = \{9,5,2\}\cap\{6\}=\phi$
 \item[3(a):] $Pin(\{(1,3)\})\cap Pin(N_4(4,1)\setminus \{(4,1)\}) = \{4\}\cap\{5,8,7\}=\phi$
 \item[3(b):] $Pin(N_4(1,2)\setminus \{(1,3)\}) \cap Pin(\{(4,1)\}) = \{10,7\}\cap\{6\}=\phi$
\end{itemize}
}
\noindent
Similar checking is required for another static droplet on $(5,4)$. It is interesting to notice that if
$m([1,2]\rightarrow [2,2])$ is executed instead of $m([1,2]\rightarrow [1,3])$, it violates the design constraint,
since $Pin(\{(2,2)\})\cap Pin(N_4(4,1)) = \{7\}\cap\{5,8,7\}\neq \phi$. \hfill $\blacksquare$
\end{example}

\subsection*{Mixing operation -- $mix([r_1,c_1]\leftrightarrow[r_2,c_2],t_{mix},mtype$):}
\noindent
Correctness checking of a mixing operation depends on the type of the mixer used since different mixers have different
sizes and mixing times. Design methodologies for assay specific pin-constrained DMFB~\cite{XuPinConstrained} assign pins
for independent mixer units in such a manner so that mixers can operate in parallel.  However, assay-independent pin
assignments require correctness checking of the mixing operation at each time step. Note that one can easily split the
mixing operation into a sequence of move operations over time. Again it depends on the type of mixer and running
time. For simplicity, we have assumed that mixers are instantiated at particular locations on the DMFB and their pin
assignments are independent.

\section{The complete verification procedure }
\label{sec:complete}
\label{subsec:eqiv}
\noindent
We will now illustrate the design constraint verification process on an example of  diluting a sample using
{\em twoWayMix}~\cite{MitraATSP}. The symbolic form of the synthesized output (actuation sequences) is also shown in 
Fig.~\ref{fig:synopsis}. The snapshot of the biochip at different time instants is shown in Fig.~\ref{fig:example}. The 
design constraint analysis process reads the symbolic fluidic instructions and verifies the operations to be executed at 
time $t+1$ (if any), depending upon the biochip description at $t$, i.e., $B_{r\times c}^t$.

Let us assume that we have set up $T_{reservior}$ from architectural descriptions correctly, and initialized the biochip
status at $t=0$, i.e.,  $B_{r\times c}^0 (=\wedge_{(\forall i,j)}\neg x_{i,j}^0)$ and set $T_{mixer}=\phi$. In the
next time step $(t+1)$, $T_{mixer}$ is checked for any expired mixer. If yes, we need to take
an appropriate action described in the $mix$ operation of Section~\ref{sec:mix}. Next, the  fluidic operations
$(I_{t+1})$, which are to be executed at $t+1$  are taken care of. Each instruction is verified as stated earlier 
and $B_{r\times c}^{t+1}$ is updated gradually. Finally, when all the executable instructions at $t+1$ are verified, 
$B_{r\times c}^{t+1}$ is evaluated for truth value. If it is \textit{true}, we proceed to the next time step and 
continue this process until all actuation sequences are processed. If any fluidic instruction executed at any instant
violates any design constraint, the potential cause is notified for debugging purpose.

\subsection*{Correctness of synthesized sequencing graph}
\noindent
The synthesized bioassay, which is verified to have no design errors, might still
suffer from certain realization 
errors. Hence, we explicitly construct the synthesized sequencing graph with the design constraint analysis process. 
Once we reconstruct a complete sequencing graph on successful completion of the design analysis phase, it is compared 
with the input graph for detecting realization errors, if any. The outline of the incremental construction of 
synthesized sequencing graph is as follows.

We assign an unique $id$ to each droplet, input reagent, output and waste reservoir and graph $G_s=(V_s,E_s)$ is 
created, where $V_s$ contains the nodes labeled with reservoir $id$s and $E_s=\phi$. Any droplet dispensed from a 
reservoir inherits the same $id$ as that of the dispensing reservoir and retains it while movement during its lifetime. 
The mix operation is handled in a different fashion. Note that it requires the insertion of an 
entry in $T_{mixer}$, which contains the $id$s of two input droplets along with start ($t_s$) and end ($t_e$) time of 
mixing operation. When a mixing operation is completed, two new droplets are created. A unique $id_{new}$ is generated 
and assigned to both of these droplets. A new node $v_{new}$ labeled with $id_{new}$ is added to $V_s$, and two edges 
$(v_1,v_{new})$ and $(v_2,v_{new})$ are added to $E_s$, where $id$s of $v_1$ and $v_2$ are the $id$s of droplets that 
were mixed together. Additionally, we store the start and end times of mixing operation in $v_{new}$. Similarly, when a 
droplet is transported to an output or to a waste reservoir, an edge is created to indicate the dispense operation. The 
following example illustrates the generation of synthesized sequencing graph with design constraint analysis.

\begin{figure*}[!ht]
\centering
\includegraphics[width=1\textwidth]{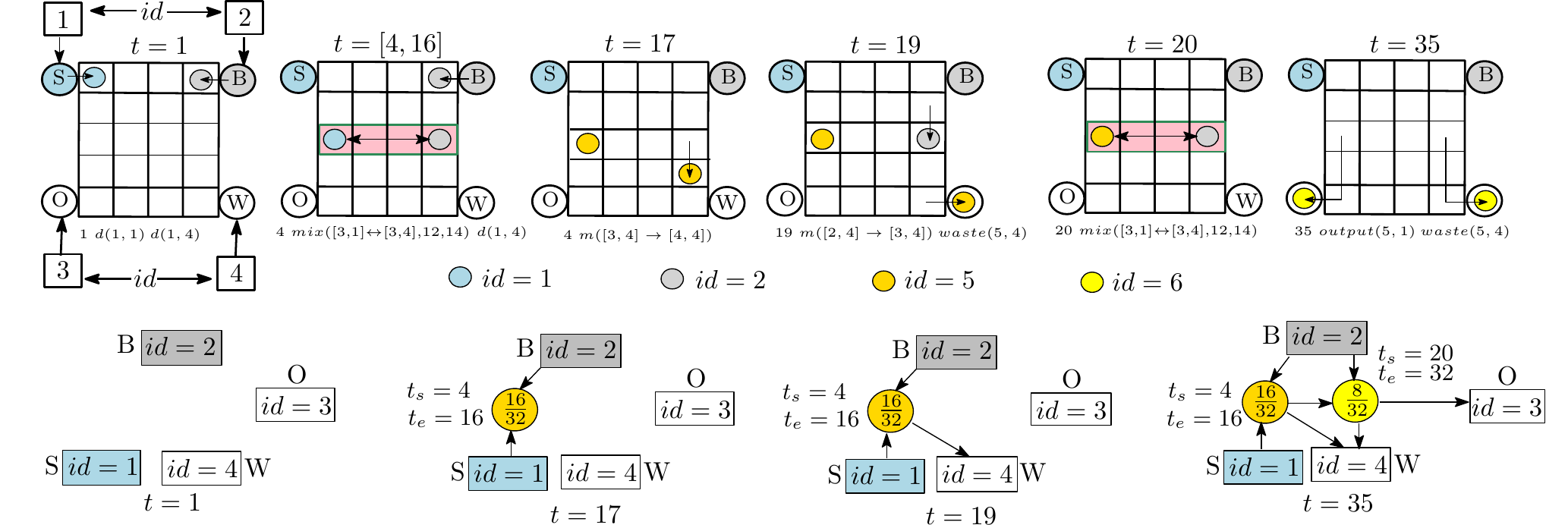}
\caption{Incremental construction of the synthesized sequencing graph}
\label{fig:constuct_syn_seq_graph}
\end{figure*}

\begin{example}
\noindent
\normalfont
Consider the bioassay as shown in Fig.~\ref{fig:synopsis}. The snapshots of the layout and the sequencing graph, which 
is incrementally reconstructed over different time steps, are shown in Fig.~\ref{fig:constuct_syn_seq_graph}. Initially, 
we assign unique $id$s to the sample and buffer dispensers, output and waste reservoirs as $id$ 1 and 2, 3 and 4 
respectively. A directed graph $G_s = (V_s,E_s)$, where $V_s = \{ S, B, O, W\}$ and $E_s = \phi$, is created.  At $t=0$, 
two droplets are dispensed from the sample and buffer reservoirs respectively. A mixer is 
instantiated with two droplets i.e., sample $(id=1)$ and buffer $(id=2)$, at $t=4$ which remains active until $t=17$.  
At $t=17$, two new droplets are created by mixing sample with buffer and a new $id\:(5)$ is assigned to each of the 
newly generated droplets. Moreover, a new node with concentration factor $\frac{16}{32}$ is added along with two edges 
i.e., $V_s = V_s \cup \{v_1\}$ and $E_s = E_s \cup \{(S,v_1),(B,v_1)\}$. Note that the newly created node $v_1$ is 
annotated with mixing start time ($t_s$) and end time ($t_e$). At $t=19$, the droplet with $id=5$ is moved to the waste 
reservoir, which in effect adds a new edge to $E_s$ i.e., $E_s = E_s \cup \{(v_1,W)\}$. Another mixer, which is 
instantiated at $t=20$, can be handled analogously. The synthesized sequencing graph is generated at $t=35$ as shown in 
Fig.~\ref{fig:constuct_syn_seq_graph}. \hfill $\blacksquare$
\end{example}

\noindent
\x{After the synthesized sequencing graph is generated and annotated with concentration factors/ratios, its conformance
with the input sequencing graph is easy to check. Two dummy nodes $v_{start}$ and $v_{end}$ are added to make each of 
them a single-entry and a single-exit graph. Since the start and end times of each mix operation are available from the 
corresponding node, it is easy to compute the mixing time. Also, the concentration factors/ratios of the droplets are 
known, and therefore, a breadth-first traversal on the two graphs is enough for checking the correspondence between 
them. However, if there was a realization error (e.g., because of a wrong mix-split operation), then a different 
concentration factor/ratio would have been generated. Hence, the proposed graph traversal can identify such unintended 
mixing operations.}

\section{Implementation}
\label{sec:implementation}
\begin{figure}[!h]
\centering
\includegraphics[width = .5\textwidth]{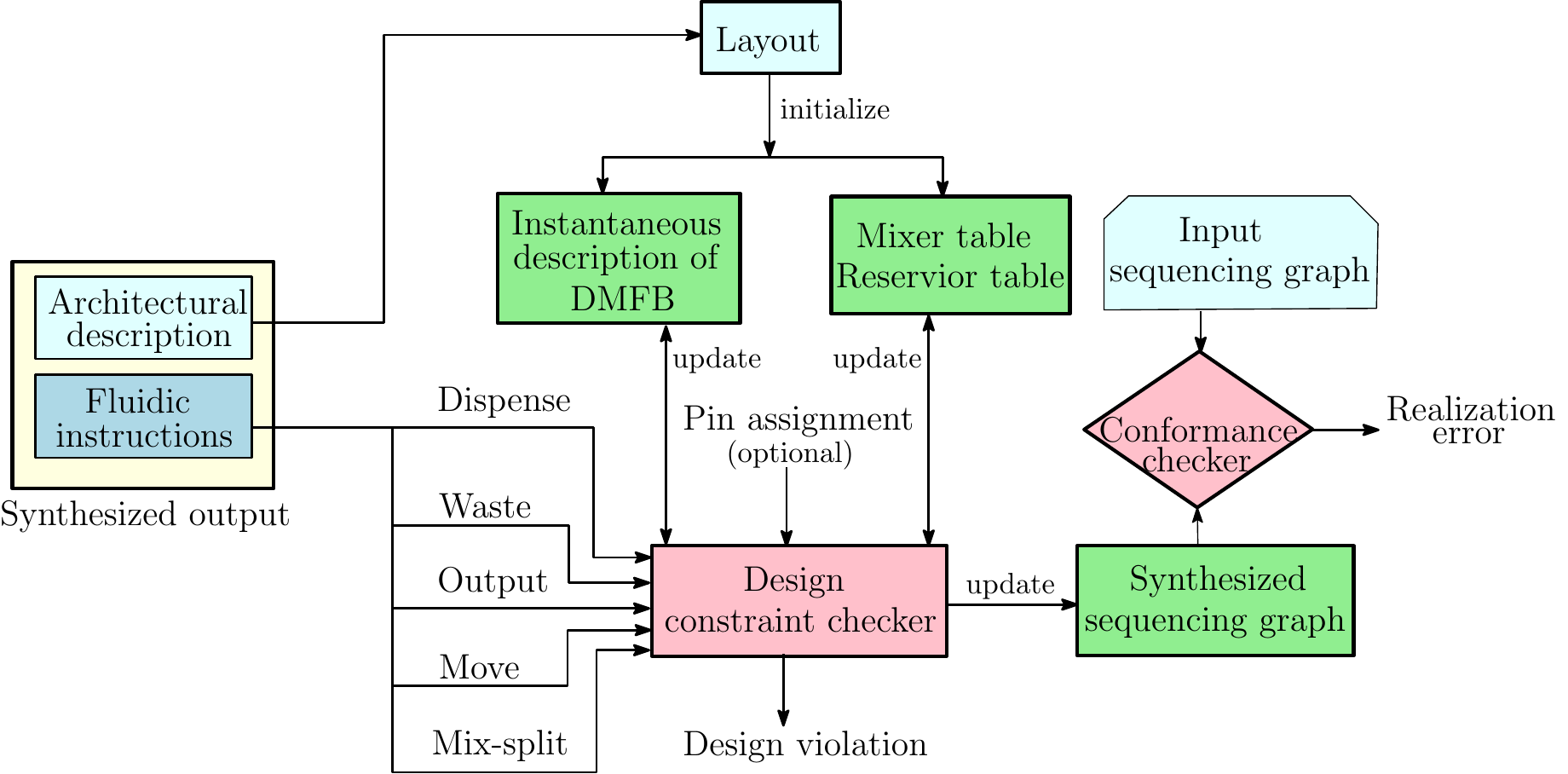}
\caption{Architecture of the verification engine}
\label{fig:implementation} 
\end{figure}

\noindent
We have developed a DMF biochip simulator (\textit{SimBioSys}) and integrated with the verification 
engine for visual simulation of a complete assay operation. The architecture of our verification tool is shown in 
Fig.~\ref{fig:implementation}. It has two major components, namely, a design-constraint checker and a conformance 
checker.

The design-constraint checker takes the symbolic representation of the synthesized output as one of the inputs and
extracts the architectural layout parameters (dimension, reservoirs) on which the synthesized bioassay will be 
executed. After extracting the DMFB layout, it initializes the reservoir table and the associated data structure for 
maintaining the instantaneous description of the DMFB with no droplet present at time $t=0$. Similarly, the mixer table 
is initialized. Next, the design-constraint checker reads the fluidic instructions to be executed 
at time $t+1$ and groups them into several fluidic instruction categories as shown in Fig.~\ref{fig:implementation}. 
Note that the fluidic instructions executed at time $t+1$ are concurrent. The verification engine checks the correctness 
of instructions in each group and generates the potential cause of design rule violation along with the fluidic 
operation that might have caused the violation.

Note that the Boolean expressions that capture the design constraints  described in Section~\ref{sec:mix} are simple in 
nature i.e., each CNF clause consists of only one variable. Hence, truth-value checking  can be accomplished without 
invoking any sophisticated satisfiability (SAT) solver. \x{We need a constraint solver that can take in a 
Boolean formula and a valuation to the variables, 
and determine if the formula evaluates to $true$ or $false$. This is more of verifying a witness on a Boolean formula 
(polynomial time), than generating a witness (computationally hard).} We have used a table-driven 
method that determines the truth value of Boolean expressions in  $O(r\times c)$ time, i.e., linear in the number of 
electrodes in the biochip. In the  case of the  pin-constrained biochip verification process, the design-constraint 
checker takes the pin assignment separately and checks the pin constraint specific design rules which are described in 
Section~\ref{sec:pin_constrained}. Additionally, it incrementally builds the annotated synthesized sequencing graph. 
Finally, conformance checking between the synthesized sequencing graph and input sequencing graph is performed. If any 
realization error is detected by the checker, it is notified to the user along with the potential cause of violation.

\section{Case studies}
\label{sec:case_studies}
\noindent
We have tested our framework on two real-life biochemical assays, namely, the polymerase chain reaction (PCR)
protocol~\cite{RoyDAC14Streaming} and the multiplexed bioassay~\cite{XuPinConstrained}. A general-purpose DMFB is used 
for verifying PCR whereas a pin-constrained DMFB platform is used for correctness checking of the multiplexed bioassay. 
In our experiments, we injected several design and realization errors in the synthesized output and observed the 
response of the tool.

\subsection*{PCR on a general-purpose DMFB}

\begin{figure*}[!t]
\centering
\begin{subfigure}{.48\textwidth}
  \centering
  \includegraphics[scale=.45]{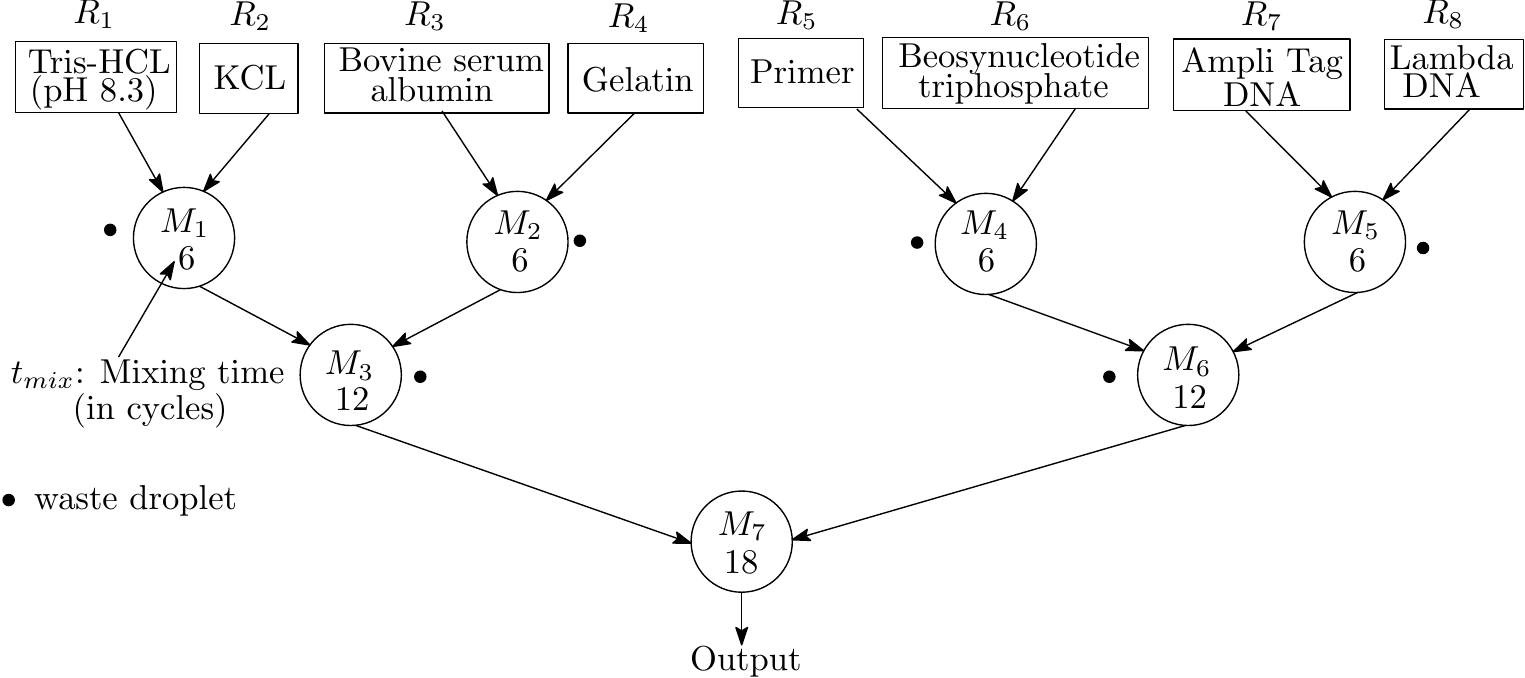}
  \caption{Mixing stage of PCR}
  \label{fig:PCR}
\end{subfigure}
\begin{subfigure}{.45\textwidth}
  \centering
  \includegraphics[scale=.55]{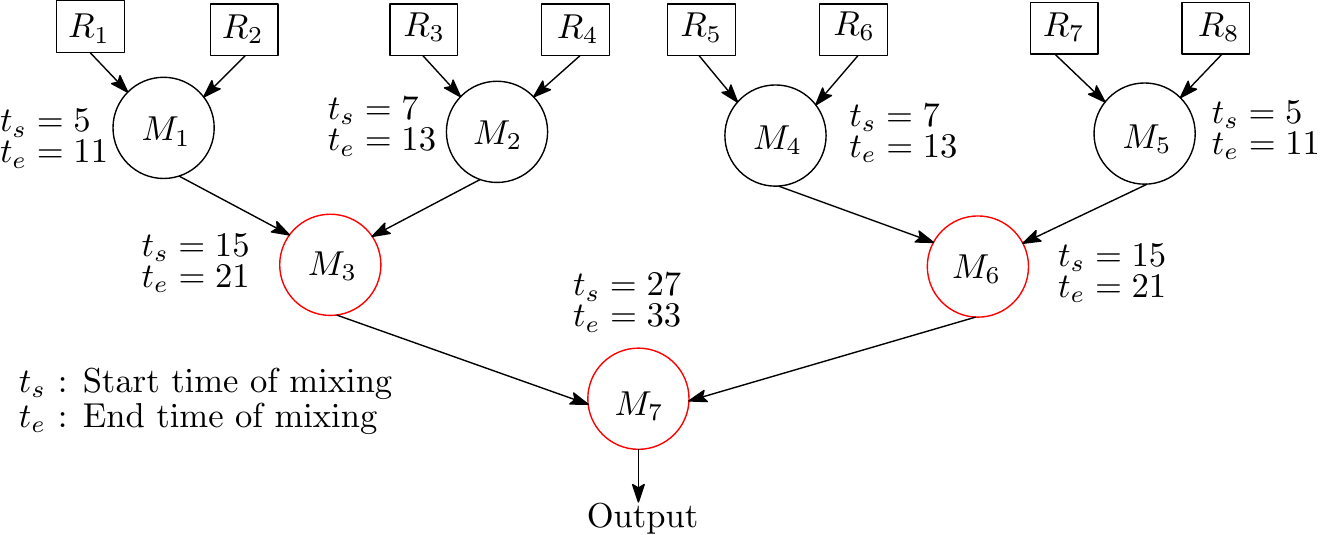}
  \caption{Synthesized PCR mixing tree}
  \label{fig:PCR_syn}
\end{subfigure}
\caption{Behavioral description of bioassay.}
\label{fig:ip-and-synthesized_PCR}
\end{figure*}

\begin{table*}
\centering
\tiny
\caption{ Symbolic representation of the actuation sequence of PCR }
\label{tab:actuation_seq}

\begin{tabular}{|p{5pt}|p{470pt}|}
\hline
\multicolumn{2}{|c|}{\textbf{Architectural layout}}\\ \hline
\multicolumn{2}{|c|}{$dim(15,15)$  $accuracy\:\:5$}\\
\multicolumn{2}{|c|}{$R(3,1,R_1)\:R(8,1,R_2)\:R(13,1,R_3)\:R(15,8,R_4)\:R(1,8,R_5)\:R(3,15,R_6)\:R(8,15,R_7)
\:R(13,15,R_8)$}\\ \hline
\multicolumn{1}{|l|}{\textbf{$t$}} & \multicolumn{1}{c|}{\textbf{Symbolic instructions executed}} \\ \hline
1 & $d(3,1)$ $d(8,1)$ $d(13,1)$ $d(15,8)$ $d(13,15)$ $d(8,15)$ $d(3,15)$ $d(1,8)$\\
2 & $m([3,1]\to [3,2])$ $m([8,1]\to [8,2])$ $m([13,1]\to [13,2])$ $m([15,8]\to [14,8])$ $ m([13,15]\to [13,14])$
$m([8,15]\to [8,14])$ $m([3,15]\to [3,14])$ $ m([1,8]\to [2,8])$\\
3 & $m([3,2]\to [3,3])$ $ m([8,2]\to [8,3])$ $ m([13,2]\to [13,3])$ $m([14,8]\to [13,8]) m([13,14]\to [13,13])$
$m([8,14]\to [8,13])$ $m([3,14]\to [3,13])$ $ m([2,8]\to [3,8])$\\
4 & $m([3,3]\to [4,3])$ $ m([8,3]\to [7,3])$ $ m([13,3]\to [12,3])$ $m([13,8]\to [12,8]) m([13,13]\to [12,13])$
$m([8,13]\to [9,13])$ $m([3,13]\to [4,13])$ $ m([3,8]\to [4,8])$ \\
5 & $mix([4,3]\leftrightarrow [7,3],6,41)$ $ mix([9,13]\leftrightarrow [12,13],6,41)$ $m([12,3]\to [11,3])$
$m([12,8]\to [11,8])$ $m([4,13]\to [5,13])$ $ m([4,8]\to [5,8])$\\
6 & $m([11,3]\to [11,4])$ $ m([11,8]\to [11,7])$ $m([5,13]\to [5,12])$ $ m([5,8]\to [5,9])$\\
7 & $mix([11,4]\leftrightarrow [11,7],6,14)$ $ mix([5,12]\leftrightarrow [5,9],6,14)$\\
14 & $m([4,3]\to [3,3])$ $ m([7,3]\to [8,3])$ $ m([11,4]\to [11,3])$ $m([11,7]\to [11,8]) m([12,13]\to [13,13])$
$m([9,13]\to [8,13])$ $m([5,12]\to [5,13])$ $ m([5,9]\to [5,8])$\\
15 & $mix([8,3]\leftrightarrow [11,3],6,41)$ $ mix([8,13]\leftrightarrow [5,13],6,41)$ $m([5,8]\to [4,8])$
$m([11,8]\to [12,8])$\\
22 & $m([8,3]\to [7,3])$ $ m([11,3]\to [11,4])$ $ m([12,8]\to [13,8])$ $m([8,13]\to [9,13]) m([5,13]\to [5,12])$
$m([4,8]\to [3,8])$\\
23 & $m([7,3]\to [6,3])$ $ m([11,4]\to [11,5])$ $m([9,13]\to [10,13])$ $ m([5,12]\to [5,11])$ \\
24 & $m([6,3]\to [5,3])$ $ m([11,5]\to [11,6])$ $m([10,13]\to [11,13])$ $ m([5,11]\to [5,10])$ \\
25 & $m([5,3]\to [5,4])$ $ m([11,6]\to [11,7])$ $m([11,13]\to [11,12])$ $ m([5,10]\to [5,9])$ \\
26 & $m([5,4]\to [5,5])$ $ m([11,7]\to [11,8])$ $m([11,12]\to [11,11])$ $ m([5,9]\to [5,8])$ \\
27 & $mix([5,5]\leftrightarrow [5,8],6,14)$\\
34 & $end$\\ \hline
\end{tabular}
\end{table*}

\begin{figure*}[!t]
\centering
\begin{subfigure}{.23\textwidth}
	\centering
	\includegraphics[scale=.4]{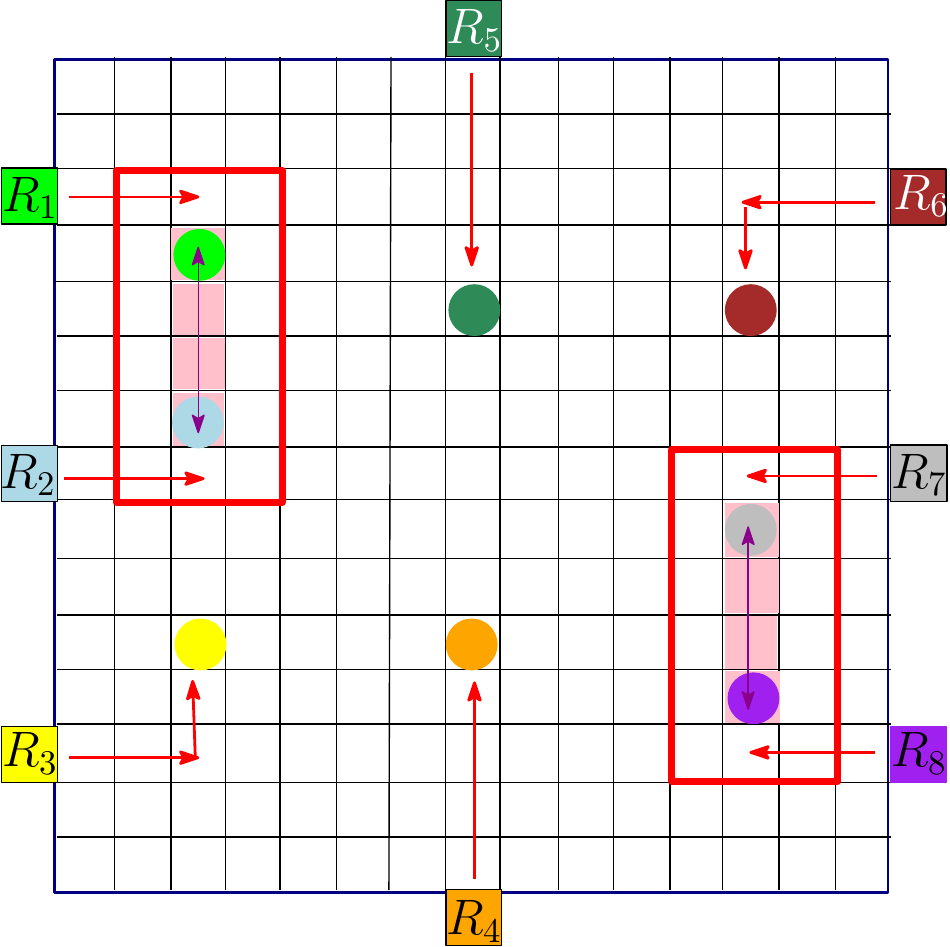}
	\caption{$t=[1,5]$}
	\label{fig:PCR_1}
\end{subfigure}
\begin{subfigure}{.23\textwidth}
	\centering
	\includegraphics[scale=.4]{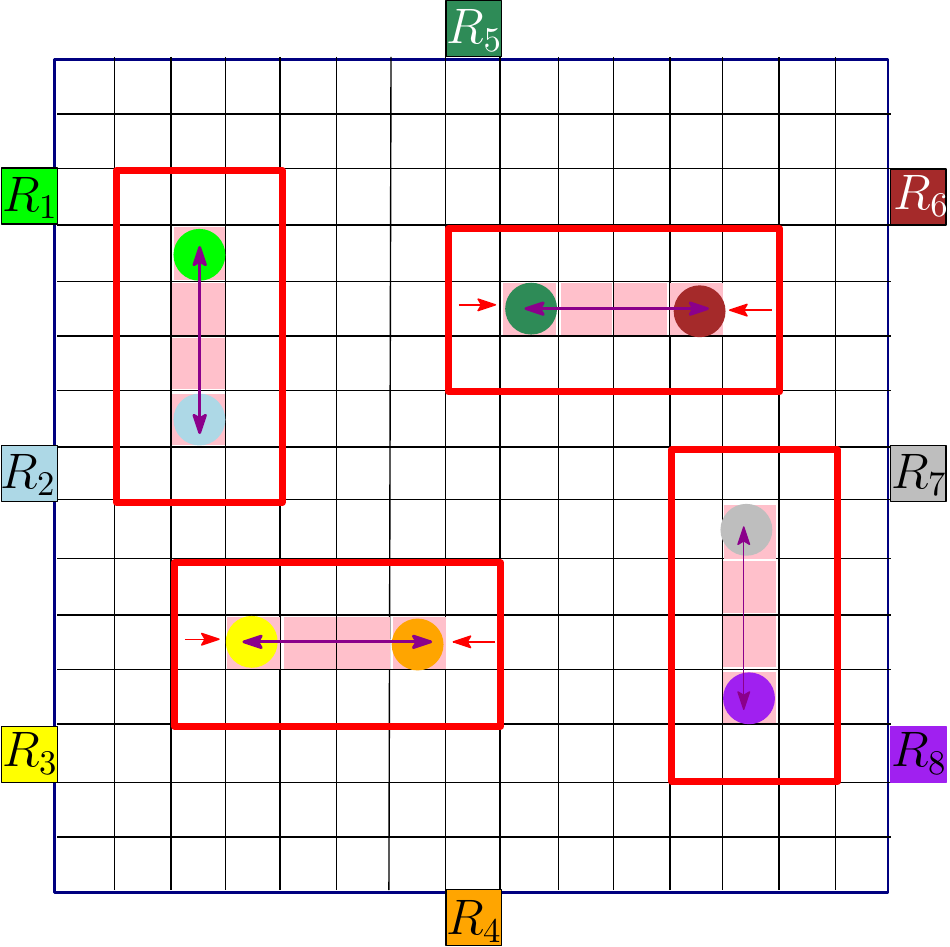}
	\caption{$t=[6,7]$}
	\label{fig:PCR_2}
\end{subfigure}
\begin{subfigure}{.23\textwidth}
	\centering
	\includegraphics[scale=.4]{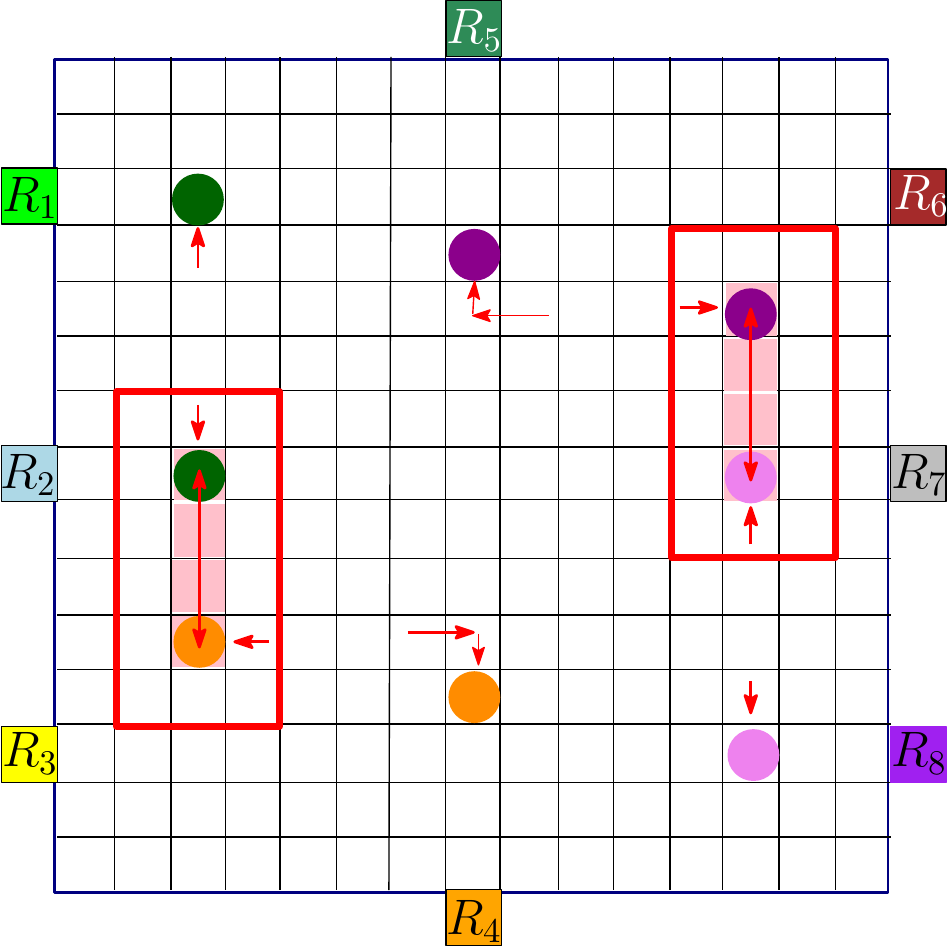}
	\caption{$t=[14,15]$}
	\label{fig:PCR_3}
\end{subfigure}
\begin{subfigure}{.23\textwidth}
	\centering
	\includegraphics[scale=.4]{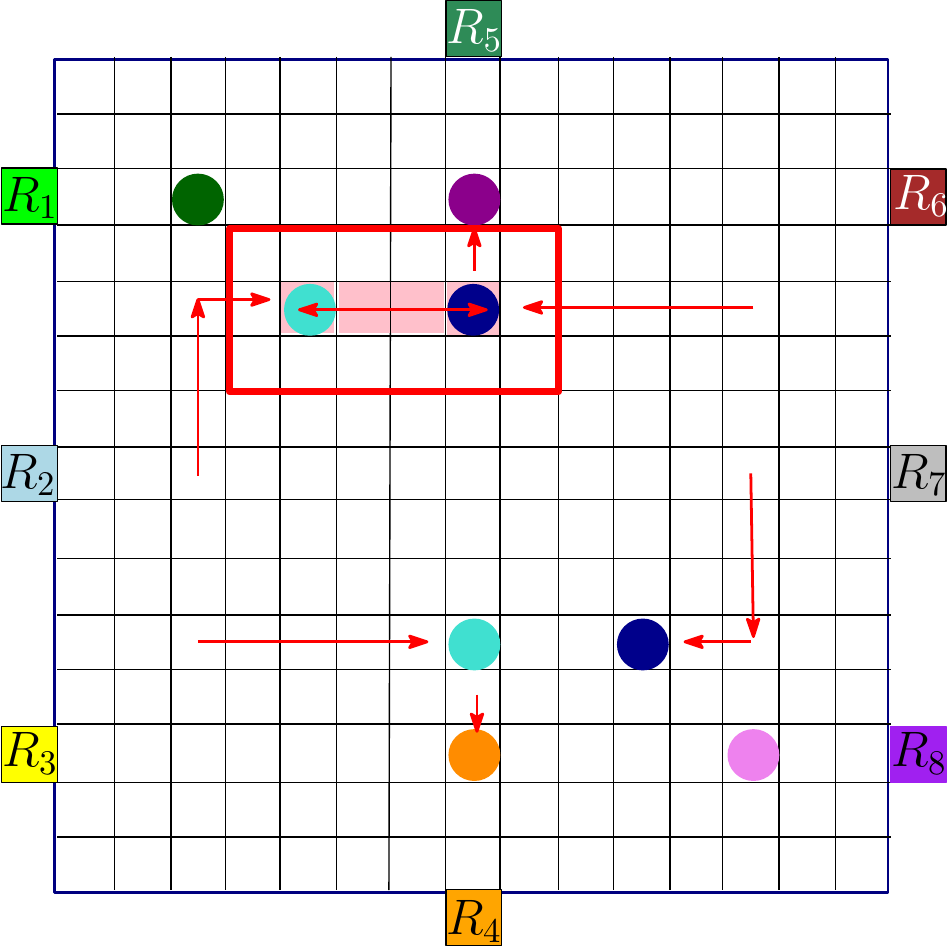}
	\caption{$t=[22,27]$}
	\label{fig:PCR_4}
\end{subfigure}
\caption{Snapshot of the execution of synthesized PCR mixing tree}
\label{fig:PCR_snapshot}
\end{figure*}

\noindent
The polymerase chain reaction is one of the most common techniques for rapid enzymatic amplification of specific DNA
fragments using temperature cycles~\cite{RoyDAC14Streaming}. Here, we use the mixing stage of PCR as an example to 
evaluate our verification tool. The sequencing graph~\cite{SuKrishTODAES04ModulePlacement} of the PCR mixing steps is 
shown in Fig.~\ref{fig:ip-and-synthesized_PCR}(a) and the symbolic representation of the synthesized
outputs~\cite{SuKrishTODAES04ModulePlacement} is shown in Table~\ref{tab:actuation_seq}. Fig.~\ref{fig:PCR_snapshot}
shows the DMFB description at different time instants along with droplet paths and mixer instantiations.


\noindent
A snapshot of \textit{SimBioSys} simulating the PCR protocol can be found in~\cite{SukantaVLSID14} . Any 
design/realization constraint violation can be easily localized by running \textit{SimBioSys}. Some results on the PCR 
protocol are reported in Table~\ref{tab:test-results-pcr}. Several design errors ($e_i$), listed in 
Table~\ref{tab:design-error}, are injected into the symbolic form of the synthesized output. The corresponding  
responses of the verification tool are listed in Table~\ref{tab:test-results-pcr}. In each case, a potential cause was 
identified correctly, and the corresponding fluidic violations were properly listed.

\ctable[
caption = Experimental results for erroneous PCR mixing,
label = tab:test-results-pcr,
pos = !h,
width = 0.5\textwidth,
framesep=0pt,
framerule = 1pt,
doinside = \scriptsize
]{p{5pt}p{120pt}p{1pt}p{73pt}}{
}{
\FL
\multicolumn{4}{c}{\textbf{Phase I - Design constraint checking }}\ML
\multicolumn{2}{c}{\textbf{Response of {\em SimBioSys}}} &
\multicolumn{1}{c}{\textbf{t}}&\multicolumn{1}{c}{\textbf{Instruction causing violation}} \ML
$e_1$ & Static fluidic constraint violated & 27& $m(11,8,12,8)$ \NN
$e_2$ & Dynamic fluidic constraint violated & 26& $m(11,7,11,8)$ $m(13,8,12,8)$ \NN
$e_3$ & Dispense from invalid input reservoir & 1& $d(2,1)$ \NN
$e_4$ & Droplet on (5,5) is in active mixer & 28& $m(5,5,5,4)$ \NN
$e_5$ & Droplet is not present on (11,4) and (11,7) & 6 & $mix(11,4,11,7,6,14)$ \ML
\multicolumn{4}{c}{\textbf{Phase II - Realization error checking }}\ML
\multicolumn{2}{c}{\textbf{Response of {\em SimBioSys}}} & \multicolumn{2}{c}{\textbf{Potential cause of error}} \ML
$e_6$ & Inhomogeneous mixing &\multicolumn{2}{l}{Mixing performed for lesser time} \NN
$e_7$ & Incorrect realization of input sequencing graph &\multicolumn{2}{l}{Wrong mix operation performed}
\LL
}

\subsection*{Multiplexed bioassay on a pin-constrained DMFB}
\begin{figure}[!t]
\centering
\includegraphics[scale=.53]{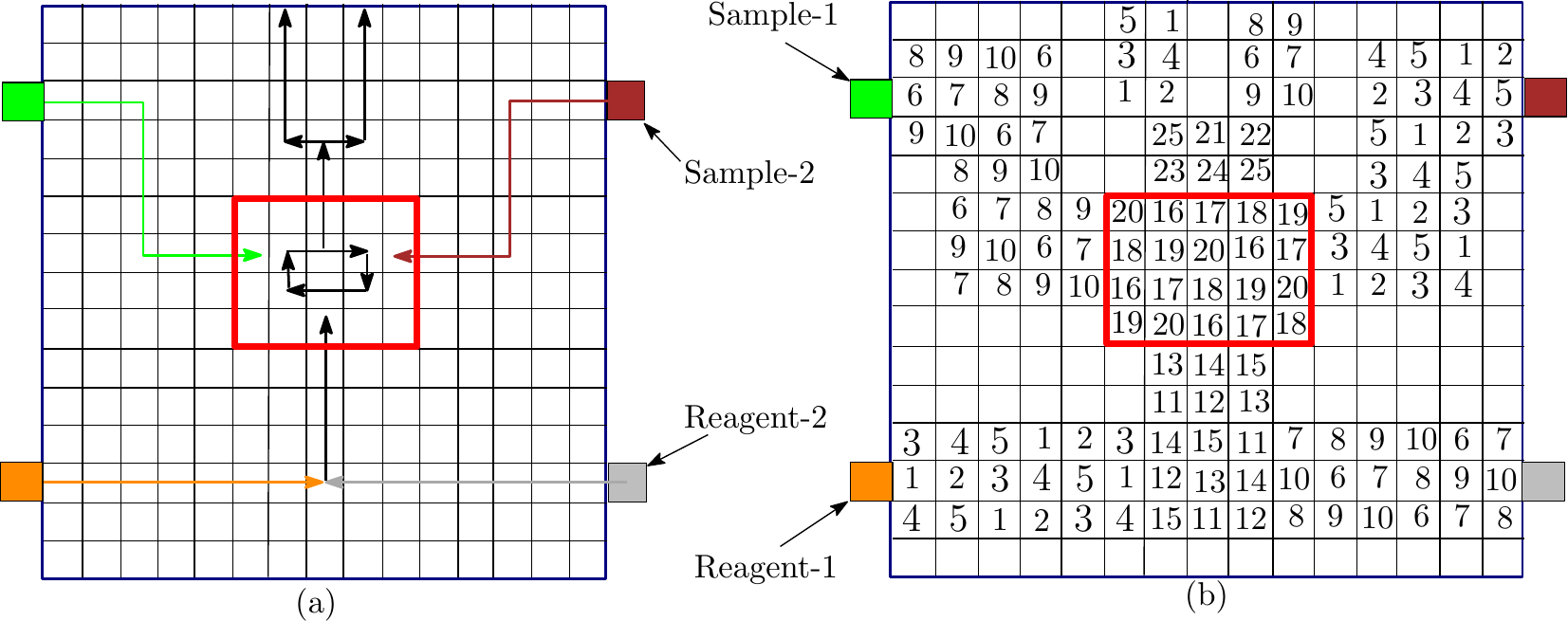}
\caption{\footnotesize Multiplexed bioassay (a) DMFB layout (b) Pin assignment}
\label{fig:mplex}
\end{figure}

\noindent
Digital microfluidic platforms have been successfully used~\cite{SistaPOC} for \textit{in-vitro} measurement of glucose
and other metabolites, such as lactate, glutamate, and pyruvate, in human physiological fluids. Hence, an
\textit{in-vitro} multiplexed bioassay for detecting human metabolic disorders has immense importance
in real life. We demonstrate our verification protocol for such a multiplexed bioassay running on a pin-constrained
DMFB. The layout and the pin assignment~\cite{XuPinConstrained} of DMFB are shown in Fig.~\ref{fig:mplex}.

We have verified the synthesized bioassay using a general purpose DMFB (dedicated pin for each electrode) in order to
ensure that the synthesized bioassay is free from any design or realization errors except those that appear from 
constrained pin assignments. Although this requirement is not mandatory, we assume it for showing only errors related 
to pin assignment. As earlier, we injected several errors into the pin assignment of the DMFB and observed the 
response of the verification engine. The corresponding  responses of the verification tool are listed in
Table~\ref{tab:test-results-mplex}. Note that in the first column, we have listed the new pin assignment of the cells
that is modified to introduce errors in the pin assignment shown in Fig.~\ref{fig:mplex}(b). In every case, the
consequences of pin assignment over fluidic instructions are detected along with the fluidic operation that
triggers the violation.

\ctable[
caption = Experimental results on pin-constrained DMFB,
label = tab:test-results-mplex,
pos = !h,
width = 0.48\textwidth,
framesep=0pt,
framerule = 1pt,
doinside = \scriptsize
]{p{73pt}p{74pt}p{1pt}p{76pt}}{
}{
\FL
\textbf{Pin assignment} &\textbf{Verifier response} & \textbf{t} &  \textbf{Instructions}\ML
$Pin(\{(13,5)\}) = 6$ & Droplet stretch & 4 & $m(3,3,4,3)$ $m(13,3,13,4)$\ML
$Pin(\{(13,15)\})=8$, $Pin(\{(14,15)\}) = 10$ & Droplet stuck on $(13,14)$ & 53 & $m(13,14,13,13)$ $m(3,14,3,13)$\ML
$Pin(\{(6,13)\}) = 7$ & Droplet stuck on $(13,11)$ & 56 & $m(5,13,6,13)$ $m(13,11,13,10)$
\LL
}

\section{Verifying fault-tolerant biochemical protocols on a cyberphysical DMFB}
\label{sec:err_recovery}
\begin{figure}[th]
\centering
\includegraphics[width=.49\textwidth]{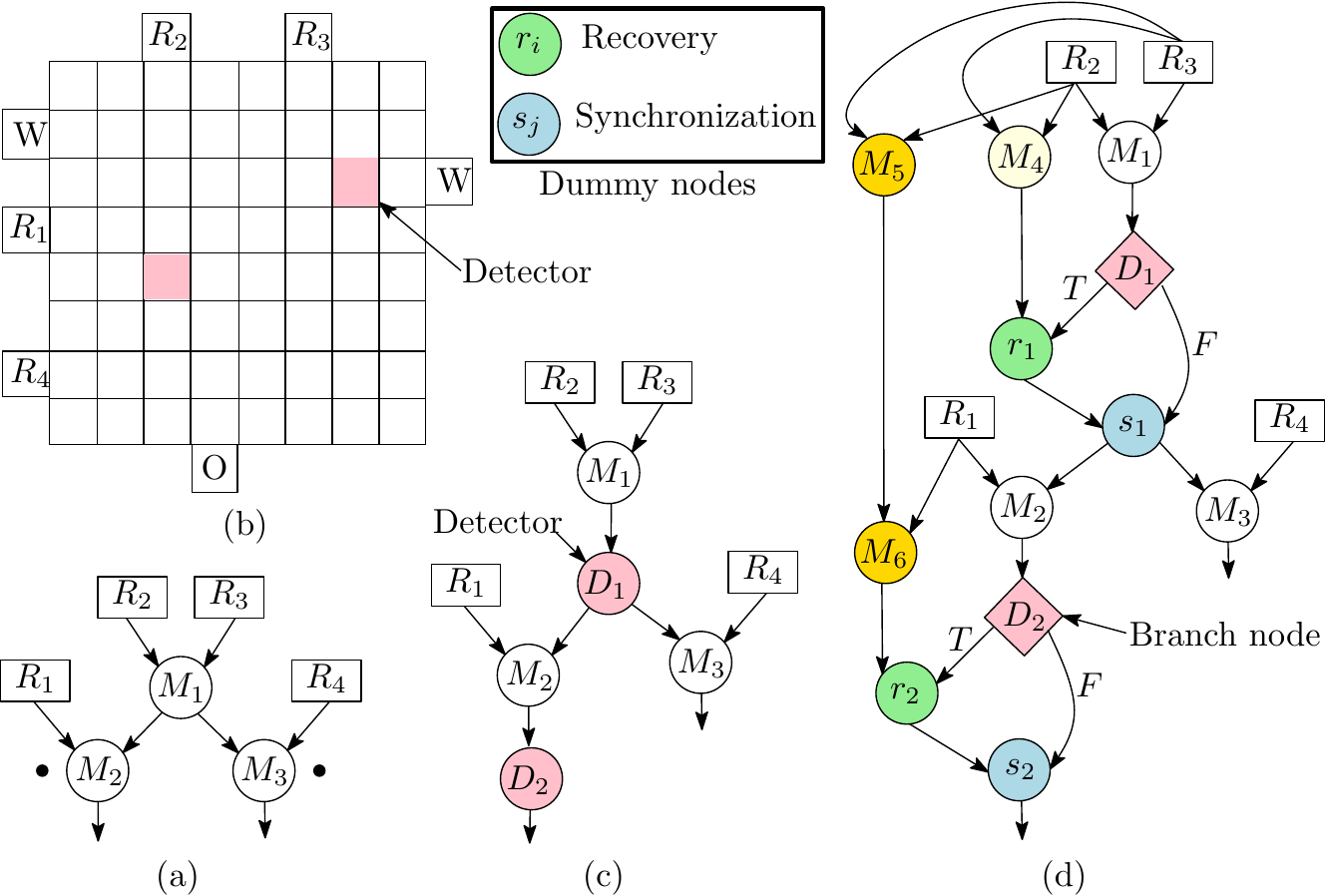}
\caption{(a) DMF platform (b) Input sequencing graph $(G)$ (c) Fault-tolerant sequencing graph graph $(G^+)$  (d) 
Sequencing graph implementing $G^+$}
\label{fig:err_recovery}
\end{figure}

\noindent
We now describe how the proposed framework can be used to verify the correctness of a fault-tolerant 
implementation of an input sequencing graph. Because of the intrinsic errors prevalent in common microfluidic 
operations, several fault-tolerant synthesis methodologies~\cite{AlisterErrorRecovery12, 
ContolPathAndErrorReovery,LuoCyberPhisicalErrorRecovery} were proposed. In such a cyberphysical system, on-chip optical 
or electronic sensors check for the presence of any volumetric/concentration error in the droplets, and their feedback 
is used to govern the assay operations in real-time. Synthesis tools for such implementations of the input sequencing 
graph $(G)$ insert detection operation for appropriate droplets in $G$. It also generates recovery operations a-priori 
for handling any detected errors. Fig.~\ref{fig:err_recovery} (a) shows the input sequencing graph $(G)$ to be 
implemented on an $8\times 8$ DMFB platform (Fig.~\ref{fig:err_recovery} (b)). A 
synthesis tool inserts two detection operations on the output droplets of mixers $M_1$ and $M_2$ in $G$, resulting in 
the sequencing graph G+ (Fig.~\ref{fig:err_recovery}(d)). We assume that no faults can occur more than once.

We need to augment our proposed instruction set (Table~\ref{tab:instructions}) with new instructions for dealing 
with error detection and recovery in a fault-tolerant implementation of $G$. The on-chip optical detectors are encoded 
as $D(d_{i},x,y,t)$, where $(x,y)$ is the location of the detector, $d_{i}$ is an unique identifier and $t$ denotes the 
number of cycles required for detection. Additionally, fluidic operation $detect(d_{i})$ starts the on-chip 
detector $d_i$, and after the detection, the corresponding flag in the DMFB controller memory is set 
appropriately~\cite{ContolPathAndErrorReovery}. Based on the error flag value of \mbox{$d_i$, $if (d_{i})\:\:\: call 
<Recovery(j)>$} 
instruction conditionally jumps to the recovery routine $Recovery(j)$, where the parameter $j$ provides a pointer to an 
appropriate error-handling routine. 

 \begin{figure}[th]
\centering
\includegraphics[width=.49\textwidth]{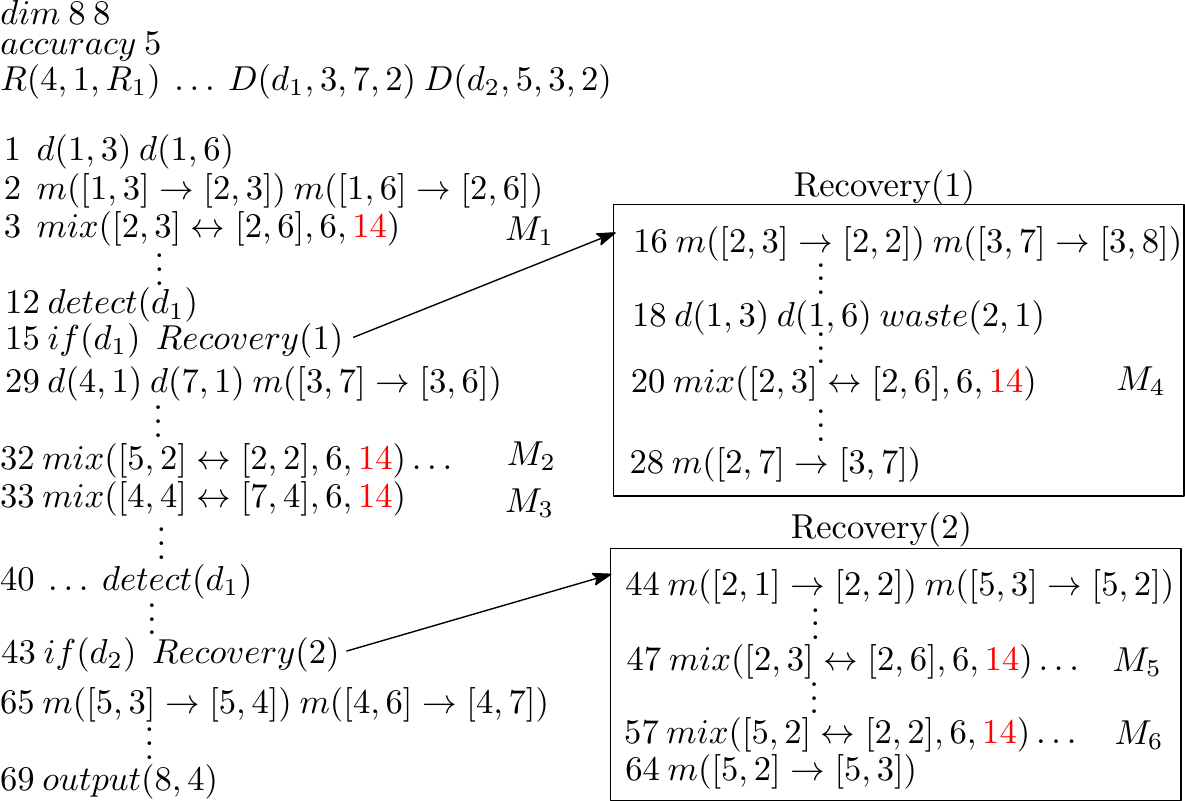}
\caption{Synthesized output with error recovery routines}
\label{fig:err_recovery_actuations}
\end{figure}
\begin{example}
\normalfont
Fig.~\ref{fig:err_recovery_actuations} shows the actuation sequence generated by the synthesis tool implementing 
Fig.~\ref{fig:err_recovery}(d) when errors are detected in the droplets generated from the mixing operations $M_1$ and 
$M_2$. Note that at $t = 15$, error recovery actions are performed ($Recovery(1)$ repeats mixing operation $M_1$ with a 
new mixing operation $M_4$); at $t = 28$, the recovery finishes. Similarly, $Recovery(2)$ instantiates two new mixing 
operations $M_5$ and $M_6$ in order to recover the error detected by detector $D_2$.\hfill $\blacksquare$ 
\end{example}
\noindent
The fluidic instructions to be executed in the synthesized bioassay depend upon the detection results.
Depending on the outcomes of the two detectors (Fig.~\ref{fig:err_recovery}(d)), there can be four possible ways the 
input bioassay can be executed. However, for verifying its correct implementation, we need to 
ensure the correctness for each of the possible execution paths. Hence, the complexity of the verification problem may 
grow exponentially in such a cyberphysical scenario.

Our verification method symbolically traverses through a valid snapshot of the DMFB at time instant $t$, 
i.e., $(B_{r\times c}^t)$ to construct another valid snapshot at $t + 1$, i.e., $(B_{r\times c}^{t+1})$, this is 
accomplished by checking the correctness of the fluidic instructions to be executed at $t$ on $(B_{r\times c}^t)$.
However, in the incremental verification process, conditional jump instructions $(if)$ may require special 
consideration. Let us consider the actuation sequence shown in Fig.~\ref{fig:err_recovery_actuations}. Note that at $t = 
15$, depending on the detection result, the  controller decides whether it needs to start an error recovery operation 
$(Recovery(1))$, or continue with normal execution (instructions following $if$). Hence, two separate streams of fluidic 
instructions can be executed at time instant $t = 16$, i.e., those from $Recovery(1)$, or those to be executed at $t = 
29$ in Fig.~\ref{fig:err_recovery_actuations}. It is important to note that Fig.~\ref{fig:err_recovery_actuations} shows 
the actuations when the droplets generated from  $M_1$ and $M_2$ are erroneous. The controller appropriately fires the 
desired instruction stream~\cite{ContolPathAndErrorReovery} depending on the detection outcome without any unnecessary 
delay. Hence at $t = 16$, two valid snapshots are possible depending on the two instruction streams. Analogously, when 
the next $if$ instruction is encountered (at $t = 43$ in Fig.~\ref{fig:err_recovery_actuations}), four separate valid 
execution paths are generated depending on the two instruction streams for each of the previous two valid snapshots. 
Thus, formally, the possible execution paths can be encoded as:
\begin{align*}
&B_{8\times 8}^0 \ldots B_{8\times 8}^{14}(B_{8\times 8}^{15} + B_{8\times 8}^{16} \ldots B_{8\times 8}^{28}) \\
&B_{8\times 8}^{29} \ldots B_{8\times 8}^{42}
(B_{8\times 8}^{43} + B_{8\times 8}^{44} \ldots B_{8\times 8}^{64}) B_{8\times 8}^{65} \ldots B_{8\times 8}^{69} 
\end{align*}
The notation $B_{8\times 8}^0 \ldots B_{8\times 8}^{14}B_{8\times 8}^{15}B_{8\times 8}^{29}\ldots B_{8\times 
8}^{42}B_{8\times 8}^{43}B_{8\times 8}^{65}\ldots \\ \ldots B_{8\times 8}^{69}$ denotes the normal execution path where 
the transition from $B_{8\times 8}^{15}$ to $B_{8\times 8}^{29}$ happens for set of fluidic operations at time instant 
29 (Fig.~\ref{fig:err_recovery_actuations}). We can thus verify all the fluidic instructions along each of the valid 
execution paths using the framework discussed in Section~\ref{sec:verification_general_purpose_dmfb}. When all such 
paths are verified for any design error, the $SG$ is reconstructed as before for the detection of 
realization errors, if any. In certain situations, it may also be possible to check the correctness of only the actual 
data path just before its execution at run-time.

\section{Conclusion and future directions}
\noindent
We have presented the foundations of a  formal correctness checking framework for post-synthesis
verification of a biochemical mixing protocol implemented on a DMF platform. Our formulation allows us to model several 
possible sources of design and implementation errors that may arise in the design of general-purpose, pin-constrained 
and cyberphysical biochips. The verification flow has been formalized and tested on the PCR protocols and 
\textit{in-vitro} multiplexed bioassay. To the best of our knowledge, such a framework does not exist in the literature. 
It paves the way  of handling several violations that may arise from imprecise synthesis strategies and change in target 
architectures. The study also opens up several new challenges concerning on-chip protocol realizations. In particular, 
the generalization of the method  for incorporating  verification of a technology-enhanced chip remains open.



\end{document}